\documentclass[aps,prx,twocolumn,preprints, superscriptaddress,amsmath,amssymb,floatfix,longbibliography]{revtex4-1}
\usepackage{tabularx}
\usepackage{ gensymb,wasysym}
\usepackage{bm}
\usepackage{epsfig,subfigure}
\usepackage{multirow}
\usepackage{graphicx}
\usepackage{color}
\usepackage{amsfonts, dsfont}
\usepackage{physics}
\usepackage{exscale}
\usepackage{amsbsy}
\graphicspath{{Fig/},{../Fig/}}
\usepackage[colorlinks,urlcolor=blue,linkcolor=blue,anchorcolor=blue,citecolor=blue]{hyperref}
\usepackage{feynmp-auto}

\newcommand{\phicf}{\phi_{\textrm{CF}}}

\newcommand{\SU}{\textrm{SU}}
\newcommand{\be}{\begin{equation}}
\newcommand{\ee}{\end{equation}}
\usepackage{slashed}
\usepackage[normalem]{ulem}

\begin{document}

\title{Emergent Multi-flavor QED3 at the Plateau Transition between Fractional Chern Insulators: Applications to graphene heterostructures}
\author{Jong Yeon Lee}
\affiliation{Department of Physics, Harvard University, Cambridge MA 02138}

\author{Chong Wang}
\affiliation{Department of Physics, Harvard University, Cambridge MA 02138}

\author{Michael P. Zaletel}
\affiliation{Department of Physics, Princeton University, Princeton, New Jersey 08540, USA}

\author{Ashvin Vishwanath}
\affiliation{Department of Physics, Harvard University, Cambridge MA 02138}

\author{Yin-Chen He}
\affiliation{Department of Physics, Harvard University, Cambridge MA 02138}
\affiliation{Perimeter Institute for Theoretical Physics, Waterloo, Ontario N2L 2Y5, Canada}
\date{\today}
\begin{abstract}
Recent experiments in graphene heterostructures have observed Chern insulators - integer and fractional Quantum Hall states made possible by a periodic substrate potential. 
Here we study theoretically that the competition between different Chern insulators, which can be tuned by the amplitude of the periodic potential, leads to a new family of quantum critical points described by QED$_3$-Chern-Simons theory.
At these critical points, $N_f$ flavors of Dirac fermions interact through an emergent U$(1)$ gauge theory at Chern-Simons level $K$, and remarkably, the \emph{entire} family (with any $N_f$ or $K$) can be realized at special values of the external magnetic field.
Transitions between particle-hole conjugate Jain states realize ``pure'' QED$_3$ in which multiple flavors of Dirac fermion interact with a Maxwell U$(1)$ gauge field.  
The multi-flavor nature of the critical point leads to an emergent $\SU(N_f)$ symmetry. Specifically, at the transition from a $\nu=$1/3 to 2/3 quantum Hall state, the emergent SU(3) symmetry predicts an octet of  charge density waves with enhanced susceptibilities, which is verified by DMRG numerical simulations on microscopic models applicable to graphene heterostructures.
We propose experiments on Chern insulators that could resolve open questions in the study of 2+1 dimensional conformal field theories and test recent duality inspired conjectures.
\end{abstract}
\maketitle
\tableofcontents

\section{Introduction} 
 Perhaps the most remarkable example of emergence in condensed matter physics is the appearance of deconfined gauge fields, which by now is well established in the context of the Fractional Quantum Hall effect, whose low energy physics is captured by a Chern Simons gauge theory. Here, we discuss a family of quantum phase transitions between different quantum Hall states that, owing to the emergent gauge fields, are strikingly different from conventional phase transitions.  

At the same time, recent experimental advances have brought their physical realization within reach.  There has been rapid  progress in achieving quantum Hall physics~\cite{Klitzing1980,Tsui1982,Laughlin1983} in the presence of a periodic potential at the scale of the magnetic length. 
Moire patterns between boron-nitride and bilayer graphene~\cite{MoireGraphene1,MoireGraphene2, MoireGraphene3} as well as patterned gate electrodes \cite{Dean2017} have been utilized to create a superlattice potential, leading to the realization of integer Chern Insulator (ICI) phases~\cite{TKKN,Haldane_honeycomb} within the  Hofstadter butterfly~\cite{Hofstadter1976}. 
While these phenomena are  present even in models of non-interacting electrons, recent experiments on bilayer graphene have demonstrated the existence of fractional Chern insulators (FCIs)~\cite{FCI_experiment}, fractional quantum Hall states that can only occur through the interplay between the periodic potential and electron-electron interactions. While there is a significant theoretical literature on FCI states \cite{Moeller2009,Neupert2011,Sheng2011, Regnault2011,Tang2011} (also see  reviews~\cite{Parameswaran2013,Bergholtz_review}), recent experimental progress calls for a study of the quantum phase transitions between the rich variety of phases which can be tuned by the periodic potential, which have attracted less attention~\cite{MacDonald_transition}.

Phase transitions between distinct quantized Hall states have been studied since the early days of the quantum Hall effect, typically in the presence of disorder which leads to plateaus of quantized Hall conductance on changing the magnetic field or electron density~\cite{Jain1990_transition, Kivelson1992_transition, Wen1993_transition, Chen1993_transition, Ye1998_transition}. 
Instead of traditional plateau transitions, we will discuss transitions that may be tuned by the amplitude of the periodic potential at fixed magnetic field and density, which cannot be realized in the disorder dominated regime~\cite{MacDonald_transition}, and have not been previously discussed.
In particular we will focus on a class of quantum phase transitions which can be approached using the composite fermion description~\cite{Jain1989,LopezFradkin1991,HLR1993}, where a new set of fermions are obtained from electrons (or bosons) by a flux attachment procedure. 
Microscopically, the periodic potential alters the dynamics of the composite fermions, inducing a change in their band topology. 
At the critical point, the composite fermions form multiple Dirac cones. 
This leads to a class of critical theories in 2+1D which take the form of  quantum electrodynamics (QED), with multiple flavors ($N_f$) of Dirac fermions coupled to a U$(1)$ gauge field arising from flux attachment. 
An additional Chern-Simons term may also be present. 
Previous discussions of the plateau transition~\cite{Jain1990_transition, Kivelson1992_transition, Wen1993_transition, Chen1993_transition} have typically considered the Chern number of the composite fermions changing by unity ($N_f=1$), in both lattice and disordered systems. This is the only natural scenario for the disordered case, and is a special case of the more general theory discussed below, where lattice translations endow the system with larger symmetry leading to several new features, in particular an enlarged $SU(N_f)$ flavor symmetry that will have important consequences. 
There have also been some studies on the FCI transition with $N_f=2$ ~\cite{Maissam2014_FQHtransition,YML2014_SPT,Tarun2013_QHtransition,Barkeshli2015}, although these apply to FCI transitions of lattice bosons rather than electronic systems.

The  family of critical theories we find are dubbed QED$_3$-Chern-Simons theory, each of which is labeled by two numbers: the number of Dirac fermion flavors $N_f$ and the Chern-Simons level $K$.
It is believed that most (if not all) of these critical theories will  flow into 2+1 dimensional conformal field theories (CFTs) in the infrared.
However the properties of those theories are rather poorly understood.
Indeed there has been a large effort to study the IR properties of  pure ($K=0$) QED$_3$ ~\cite{QEDCSB1,QED3CFT,QEDQMC1,QEDQMC2,kapustinqed,Hermele2004_Stability,Hermele2005,Hermele2005_Erratum,GroverFT,QED3RG,ChesterPufuScalarOperators,qedcft,DyerMonopoleTaxonomy,ChesterPufuBootstrappingQED}, but calculating properties at small $N_f$ remains an open issue.
Several of these theories  have applications to other long-standing problems in condensed matter physics, for example, the theory with $N_f=4$, $K=0$ has appeared in theories of high temperature superconductivity in the cuprates~\cite{RantnerWen2002} and  spin liquid physics in frustrated magnets~\cite{Hastings2000,Hermele2005,HermeleKagome,Ran2007,YCH_kagomespectrum}.
They also have interesting duality properties ~\cite{sonphcfl,wangsenthil15b,MaxAshvin15,Seiberg2016395,karchtong}, and in light of the duality proposal the self-dual $N_f=2$, $K=0$ theory \cite{tsmpaf06,qeddual,karchtong,WangDCPdual,Benini2017_BF_dual} has a surprising connection to the ``deconfined'' critical points that were first discussed in the context of quantum magnets~\cite{deccp,deccplong,lesikav04}. 

More generally, understanding the properties of  interacting CFTs is a central quest in various fields of physics. 
In 2+1 dimensions, there are a few CFTs that are well understood and can be realized experimentally, such as the Wilson-Fisher theories related to spontaneous symmetry breaking~\cite{sachdev2007quantum}.
The FCI transitions discussed here could significantly expand the list, as the whole family of  QED$_3$-Chern-Simons theory may be realizable within the experimental scenario we discuss.
For example, we show pure QED$_3$ with arbitrary $N_f$ can be realized at the transitions between the particle-hole conjugate partners of the Jain sequence states.
From an experimental point of view, the FCI transitions can be accessed by tuning the strength of periodic potential, which has already been demonstrated experimentally~\cite{Yankowitz2017, Dean2017}.
Various critical exponents may also be measurable from the charge density wave susceptibility and tunneling conductance.
We therefore believe that  future experimental study may provide new insights on the long-standing and interdisciplinary problems regarding interacting CFTs in 2+1 dimensions. 

The rest of the paper is organized as follows. In Sec.~\ref{sc:free_fermion}, we review  free fermion phase transitions between integer Chern insulators, where multiple Dirac cones can appear at special values of the magnetic field and electron density where they are protected by  magnetic translation symmetry~\cite{Langbein1969}.
The free-fermion physics is an important building block of our analysis, since a fractional Chern insulator can be understood as an integer Chern insulator of composite fermions~\cite{KolRead1993,Moeller2015}.
In Sec.~\ref{sc:FQH_FCI}, we consider a concrete example of a phase transition between two neighboring FCI phases with $\sigma_{xy} = 1/3$ and $\sigma_{xy}=2/3$. We show that the transition can be realized in an experimentally feasible model,  provide numerical confirmation using iDMRG simulations, and use the composite fermion approach to demonstrate how pure QED$_3$ arises at the critical point.  

In Sec.~\ref{sc:family}, we discuss the general composite fermion construction of FCI transitions: intuitively, they arise as Chern number changing transitions of composite fermions.
Therefore, like the free-fermion case, the magnetic translation symmetry of composite fermions can give rise to multiple Dirac cones, but with the added physics of an emergent dynamical U$(1)$ gauge field.
We show how the whole family of QED$_3$-Chern-Simons theory can  emerge at such critical points, and in particular pure QED$_3$ theories can be realized at transitions between particle-hole conjugate partners of the Jain sequence.
In Sec.~\ref{sc:physical_property}, we elaborate on the physical properties of  QED$_3$-Chern-Simons theories.
For example, a critical theory with $N_f$ Dirac fermions will have an emergent $\SU(N_f)$ flavor symmetry 
which we show physically correspond to the order parameters of the charge-density-wave at $N_f^2-1$ distinct crystal momenta. The corresponding scaling dimensions are calculated analytically within a large-$N_f$ expansion. Using iDMRG simulations, we indeed find evidence for the emergence of $\SU(3)$ symmetry at the critical point between $\sigma_{xy}=1/3$ and $\sigma_{xy}=2/3$ phases, implying the emergence of pure $N_f=3$ QED$_3$. 
We also discuss the properties of monopole operators, which  in this context are nothing but the electron creation/destruction operators and their experimental consequences for probes sensitive to the electronic  Greens function. In Sec.~\ref{sc:level-rank} we point out that the  transition of the bosonic $\nu=1/2$ Laughlin state into a Mott insulator is strongly constrained by the level-rank duality results of Benini, Hsin and Seiberg ~\cite{seiberg2,Benini2017} and discuss their physical consequences.
In Sec.~\ref{sc:experiments}, we  further discuss the experimental setup for the FCI transitions, including a more experimentally straightforward transition between a $\sigma_{xy}=1/3$ FQH state and a $\sigma_{xy}=1$ Chern insulator.
In Sec.~\ref{sc:summary}, we summarize and discuss  future interesting directions.
The appendix is devoted to the details of large-N calculation and generalizations of microscopic models for the phase transitions.

\section{Free fermions: a warm up} \label{sc:free_fermion}
We begin by reviewing transitions between integer Chern insulators at fixed density and flux, since these form the building block of our later fractionalized analysis. 
Throughout, we use units in which the unit cell of the lattice is $a^2 = 1$ and  $\frac{e^2}{\hbar} = 1$, so that the fundamental flux quantum is $\phi_0 = 2 \pi$ and the fundamental conductance is $\frac{e^2}{h} = \frac{1}{2 \pi}$. However, when we quote the value of $\sigma_{xy}$, we will express it in units of $\frac{e^2}{h}$ so that the $2\pi$ factor drops out. 
The magnetic field is measured in terms of the density of flux quanta per unit cell, $\phi = \frac{a^2 B}{\phi_0}$. 
Chern insulators are best understood through the relation between the electron density per unit cell $n$ and the flux density $\phi$: any gapped phase of matter must obey the Diophantine condition \cite{Streda,MacDonald1983}
\begin{align}
n = C \phi + s
\end{align}
where $C, s$ are invariants which characterize the phase.
From the Streda formula~\cite{Streda}, the Hall conductance is $\sigma_{xy} = \frac{d n}{d B} = C / 2 \pi$, while $s$ can be thought of as the density of electrons ``glued'' to each lattice cell~\cite{MacDonald1983}.
For any free fermion state, $C, s \in \mathbb{Z}$, while they may be rational fractions in an FCI. Galilean invariance requires $s = 0$, so non-zero $s$ indicates strong lattice effects - these are the Chern insulators.

\begin{figure}
\includegraphics[width=0.48\textwidth]{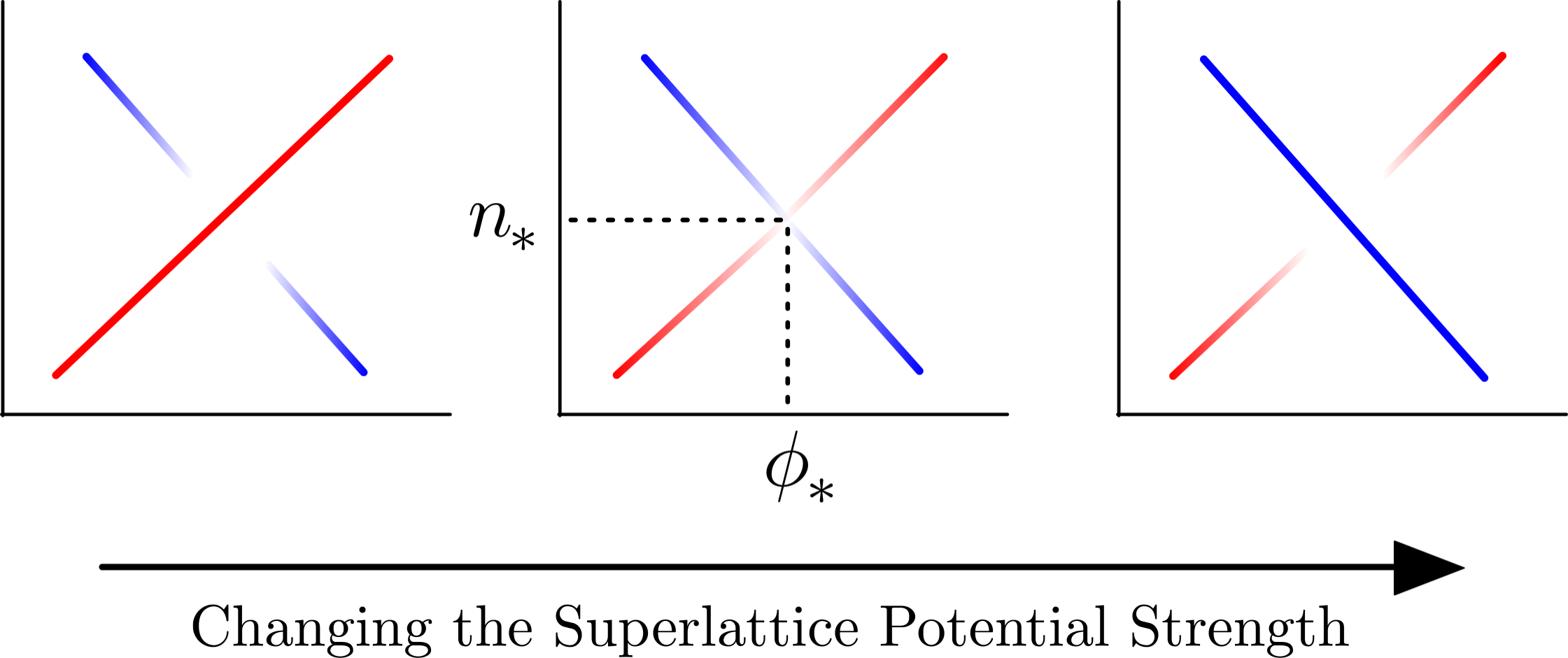} \hspace{0.1in}
\vspace{-0.1in}
\caption{\label{fig:general_transition} Red and blue lines represent gapped trajectories labeled by $(C,s)$ and $(C',s')$ respectively, which intersect at $(n_\ast, \phi_\ast)$. For a generic value of the tuning parameter $m$ (superlattice potential strength), one type of gap will ``win'' over the other at the crossing. At the critical value $m = m_{c}$, the gap closes and their relative strength is exchanged; this is the manifestation of a Chern-number changing transition.
}
\end{figure}

	Experimentally, $C$ and $s$ can be determined by measuring electrical properties such as the resistance or compressibility  in the plane of $n$ and $\phi$ (a ``Wannier plot'').
Gapped FCI and ICI phases appear as lines of incompressibility in this plane, from which the invariant can be read off as a slope and intercept (though of course the phases will only extend over a finite range of $\phi$)~\cite{MoireGraphene1,MoireGraphene2, MoireGraphene3, FCI_experiment}.
Importantly, two such lines may cross at some point $\phi_\ast, n_\ast$, indicating competition between two different ICIs or FCIs.
In Fig.~\ref{fig:general_transition}, we illustrate how this competition may evolve as a tuning parameter, such as the lattice potential, changes. The gap at $\phi_\ast, n_\ast$ closes and reopens as it transfers between the two different trajectories, indicating a Chern-number changing transition.

The Diophantine condition places a constraint on the Chern-number changing transitions which can occur at fixed density and flux.
If  $C$ and $s$ change by $\Delta C, \Delta s$ across the transition, we must have $0 = \Delta C \phi_\ast + \Delta s$. Letting $\phi_\ast = \frac{p}{q}$, this implies $p \, \Delta C  = -q \, \Delta s $. Since $p, q$ are co-prime, we conclude that $\Delta C$ is an integer multiple of $q$. 
Thus while naively one might suppose only $|\Delta C| = 1$ transitions are generic, fractional flux stabilizes higher Chern number changing transitions~\cite{Streda,MacDonald1983,Lu_enforcedHall}.

In the free fermion limit, $C$-changing transitions arise from gap-closings in the underlying band theory. The total Chern number of an integer phase is determined by summing up the Chern numbers of occupied bands. 
As a parameter such as the lattice potential amplitude changes, the conductance and valence bands may touch, closing the gap and transferring Chern number between them. The transfer of Chern number $\Delta C$ between the two is mediated by the formation of $|\Delta C|$ Dirac points, as illustrated in Fig.~\ref{fig:3Dirac}.
We review the field-theoretic description of the resulting transition to illustrate the role of the invariant $s$.
The electromagnetic response theory for a $C, s$-insulator is
\begin{align}
\label{eq:LCt}
\mathcal{L}_{\textrm{eff}}[A] = \frac{C}{4 \pi} A  dA + s A_0, 
\end{align}
where $A$ is the external electromagnetic gauge field, as verified by $\frac{\delta \mathcal{L}}{\delta A_0} = n = C \frac{B}{2 \pi} + s$.
Here (and later) $AdA$ is a short hand notation for the Chern-Simons term $A\wedge dA$.
For a transition between invariants $(C_2, s_2)$ and $(C_1, s_1)$ at rational flux $\phi_\ast$, we define a vector potential $A_\ast = (0, \textbf{A}_\ast)$ such that $\nabla \times \textbf{A}_\ast = 2 \pi \phi_\ast$ produces the uniform magnetic field associated to $\phi_\ast$, and let $\Delta C = C_2 - C_1, \bar{C} = (C_2 + C_1)/2$. The proposed Lagrangian is
\begin{equation}\label{eq:free_dirac}
\mathcal L = \sum_{I=1}^{\Delta C}  \bar \psi_I  (i\slashed{\partial}+ \slashed A -  \slashed A_\ast- m) \psi_I + \frac{\bar{C}}{4\pi} (A - A_\ast)  d(A - A_\ast) +  n_\ast A_0.
\end{equation}
Subtracting $A_\ast$ ensures that even though the  Dirac fermions couple to $A$, they see no net field at $\phi_\ast$. 
The mass $m$ is a phenomenological parameter which tunes across the transition. When $m$ is finite, a gap opens and the each Dirac fermion induces a Chern-Simons response proportional to $\frac{1}{2} \frac{m}{|m|}$. Integrating out the fermions, we obtain
\begin{align}\label{eq:free_dirac_eff}
\mathcal L &= (\bar{C}  + \frac{1}{2}  \Delta C \frac{m}{|m|} ) (A - A_\ast) d (A - A_\ast) / 4 \pi +  n_\ast A_0 \\
&= (\bar{C} + \frac{1}{2} \frac{m}{|m|} \Delta{C}) A  dA  / 4 \pi+  ( \bar{s} +  \frac{1}{2} \frac{m}{|m|} \Delta s )  A_0
\end{align}
where we have used the Diophantine relations $\Delta C \phi_\ast = - \Delta s$ and $n_\ast = \bar{C} \phi_\ast + \bar{s}$.
This ensures we produce the desired response of Eq.~\ref{eq:LCt} on either side of the transition.

This leaves open why the mass of the Dirac fermions are identical; in the absence of a symmetry relating them, the transition would instead occur through a sequence of $\Delta C=1$ transitions. The desired symmetry is of course the magnetic algebra. Letting $T_x, T_y$ denote translations by the Bravais vectors of the lattice, 
\begin{equation}
T_x T_y = e^{ 2\pi i \phi } T_y T_x
\end{equation}
Since the magnetic algebra acts projectively on the fermions, at flux $\phi = \frac{p}{q}$ this required $q$-fold degeneracies, guaranteeing the Dirac points come in groups of $\Delta C = q$.  
This can be seen explicitly in $k$-space~\cite{Langbein1969}.
Defining a magnetic Brillouin zone with respect to the commuting translations $T_x, T_y^q$, the symmetries act on eigenstates $\ket{k_x, k_y, j}$ according to 
$T_x \ket{k_x, k_y, j} = e^{i k_x} \ket{k_x, k_y, j}$ ($j$ is a band index) and 
$T_y \ket{k_x, k_y, j} = e^{i k_y} \ket{k_x+2\pi p / q , k_y, j}$.
The latter relation cyclically permutes $q$ points in the magnetic Brillouin zone.
Therefore if the system has a Dirac cone (denoted by $\psi_1$) at $(k_x, k_y)$, another $q-1$ copies will  automatically appear at $(k_x + 2\pi I  p / q, k_y)$. 
Under the translational operation the resulting $q$-flavors of Dirac fermion transform as
\begin{eqnarray}\label{mtrans}
T_y \psi_I &=& e^{i k_y} \psi_{I+1}, \quad I=0, \cdots, q - 1. \nonumber \\
T_x \psi_I &=& e^{i k_x + i 2\pi I \frac{p}{q} } \psi_{I}
\end{eqnarray}
This forms a subgroup of $\SU(q)$ under which only the singlet mass term $m\sum \bar \psi \psi$ is symmetric, which is the tuning parameter for the phase transition.
Other mass terms, i.e. $\bar \psi_I M_{IJ} \psi_J$ ($M\neq \mathds{1}$), will break the magnetic translation symmetry.

As the magnetic flux deviates away from $\phi_\ast$, the residual field is seen by the Dirac fermions.
Together they will form $\Delta C$-fold degenerate Landau levels, opening up Hall gaps at total Chern number $C = \bar{C} + (l + \frac{1}{2}) \Delta C, l \in \mathbb{Z}$.
These are precisely the trajectories allowed by the Diophantine condition $n = C \phi + s$ near $\phi_\ast$, and correspond to the formation of Chern $\Delta C$-bands. 
This makes precise the sense in which a Chern $C$-band is like ``$C$-copies'' of a Landau level; these copies are just the valley index of the $C$-Dirac fermions, which are permuted under the magnetic algebra as a subgroup of SU($C$).

\subsection{Model realization}

Chern number changing transitions can be realized in any model without Galilean invariance, such as tight-binding models or continuum Landau levels with an additional weak lattice potential.
While the former have been the subject of most numerical studies, the latter are most relevant to recent experiments in graphene heterostructures (see Sec.~\ref{sc:experiments}), so will be our focus.
In this paradigm, a strong magnetic field applied to a 2DEG first gives rise to flat  Landau levels with a magnetic length $\ell_B$ much larger than the atomic scale, split by a cyclotron energy $\hbar \omega_c$.
An additional ``superlattice'' is then applied (via lithography or a moire pattern) at a length scale comparable to $\ell_B$  but with an amplitude small compared to $\hbar \omega_c$. 
In this limit, it is justified to project the superlattice into the Landau level(s) closest to the Fermi level.
Landau level projection kills off wavevectors large compared to $\ell_B^{-1}$, so we consider superlattices of the general form
\begin{equation}\label{eq:potential}
\mu(\bm r) = U_0 \int d \bm r \sum_{m} ( V_m e^{i \bm r \cdot \bm G_m} n_{\bm r} +h.c.),
\end{equation}
where $G_m$ are the smallest several reciprocal vectors of the superlattice.
Here we consider a square lattice potential with $\bm G_1 = \frac{2\pi}{a}(1, 0)$,  $\bm G_2 = \frac{2\pi}{a}(0, 1)$, $\bm G_3 = \frac{2\pi}{a}(1, 1)$,  $\bm G_4 = \frac{2\pi}{a}(1, -1)$, and $V_1=V_2$, $V_3=V_4$, where $a$ is a lattice constant.

\begin{figure}
\includegraphics[width=0.49\textwidth]{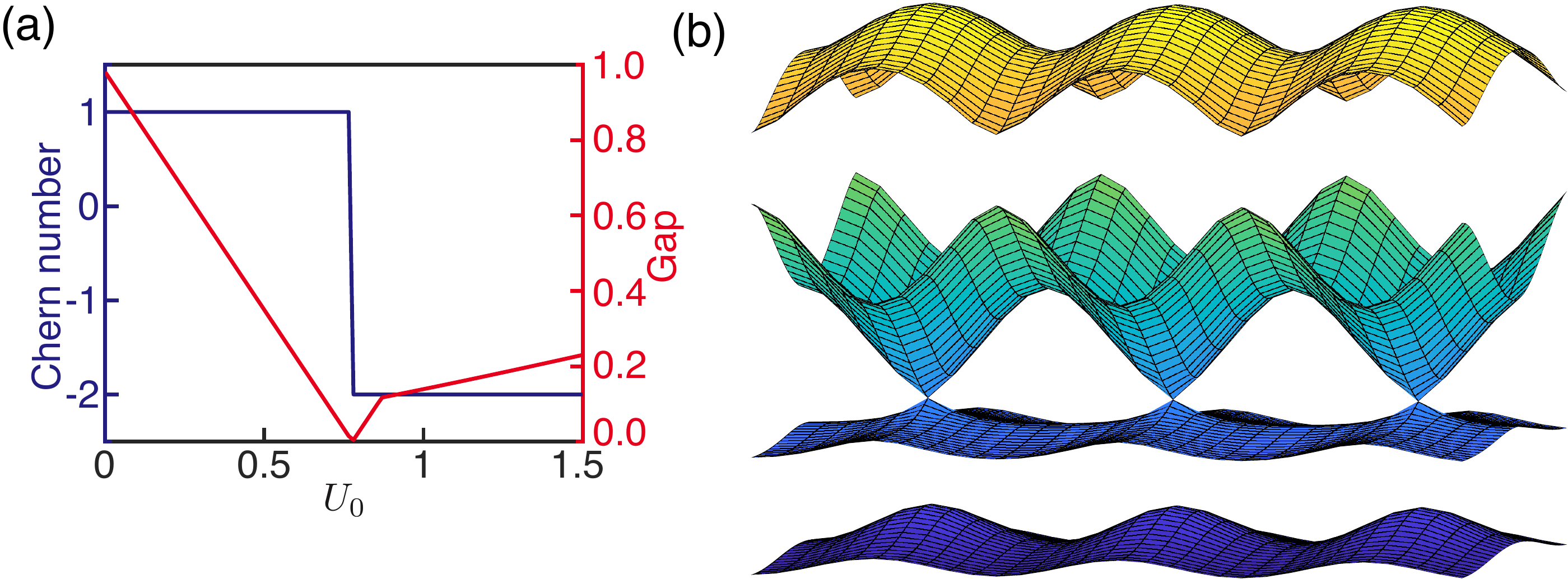}
\caption{\label{fig:3Dirac} Landau levels in the presence of a superlattice potential Eq.~\eqref{eq:potential}. We consider $\phi=2/3$,$V_1=V_2=1$, $V_3=V_4=1.4$. The superlattice is projected into the $N=1$ and $N=2$ Landau levels with spacing $\Delta E_{12}=1$. Each Landau level splits into 2 sub-bands, so that at $n = 2/3$ the 2 lowest sub-bands are filled. (a) Chern number and gap versus potential strength $U_0$ at $n = 2/3$. (b) The band structure at the transition ($U_0\approx 0.79$) shows 3 emergent Dirac cones which mediate the transition between the C=1 and C=-2 insulators.}
\end{figure}

In the presence of a finite potential the flat Landau levels broaden and split (and neighboring Landau levels may mix once the potential strength $U_0$ is comparable to the Landau level splitting).
Solving for the band structure via  standard methods~\cite{Langbein1969, Pfannkuche,usov1988theory}, we see that Chern-number changing transitions can be induced simply by tuning the potential strength $U_0$ or the lattice shape $V_{m}$.
At flux $\phi_\ast = p/q$ per unit cell of the superlattice, a Landau level splits into $p$ sub-bands each with a $q$-fold degeneracy.
As the $U_0, V_m$ change, the subbands can touch at $q$-fold Dirac points.
Fig.~\ref{fig:3Dirac} shows an example at $\phi=2/3$, where three Dirac cones appear as the total Chern number changes from $C=1$ to $C=-2$.

In addition to forming the building block of our subsequent fractional analysis, these integer-Chern number changing transitions are interesting in their own right. The multi-flavor character of the Dirac fermions will have several experimental consequences, for example, in Sec.\ref{sc:conductivity} we will discuss the electrical conductivity of these transitions.

\section{Transition between the $\sigma_{xy}=1/3$ FQH effect and a $\sigma_{xy}=2/3$ FCI}\label{sc:FQH_FCI}
In the presence of strong interactions, gaps can arise at fractional $C$ and $s$; these are the fractional quantum  Hall effects ($C \neq 0, s = 0$) and fractional Chern insulators $(C \neq 0, s \neq 0)$.
Transitions between them are the main focus of the rest of the paper.

\subsection{Model \label{sc:Model}}

\begin{figure}[t]
\includegraphics[width=0.49\textwidth]{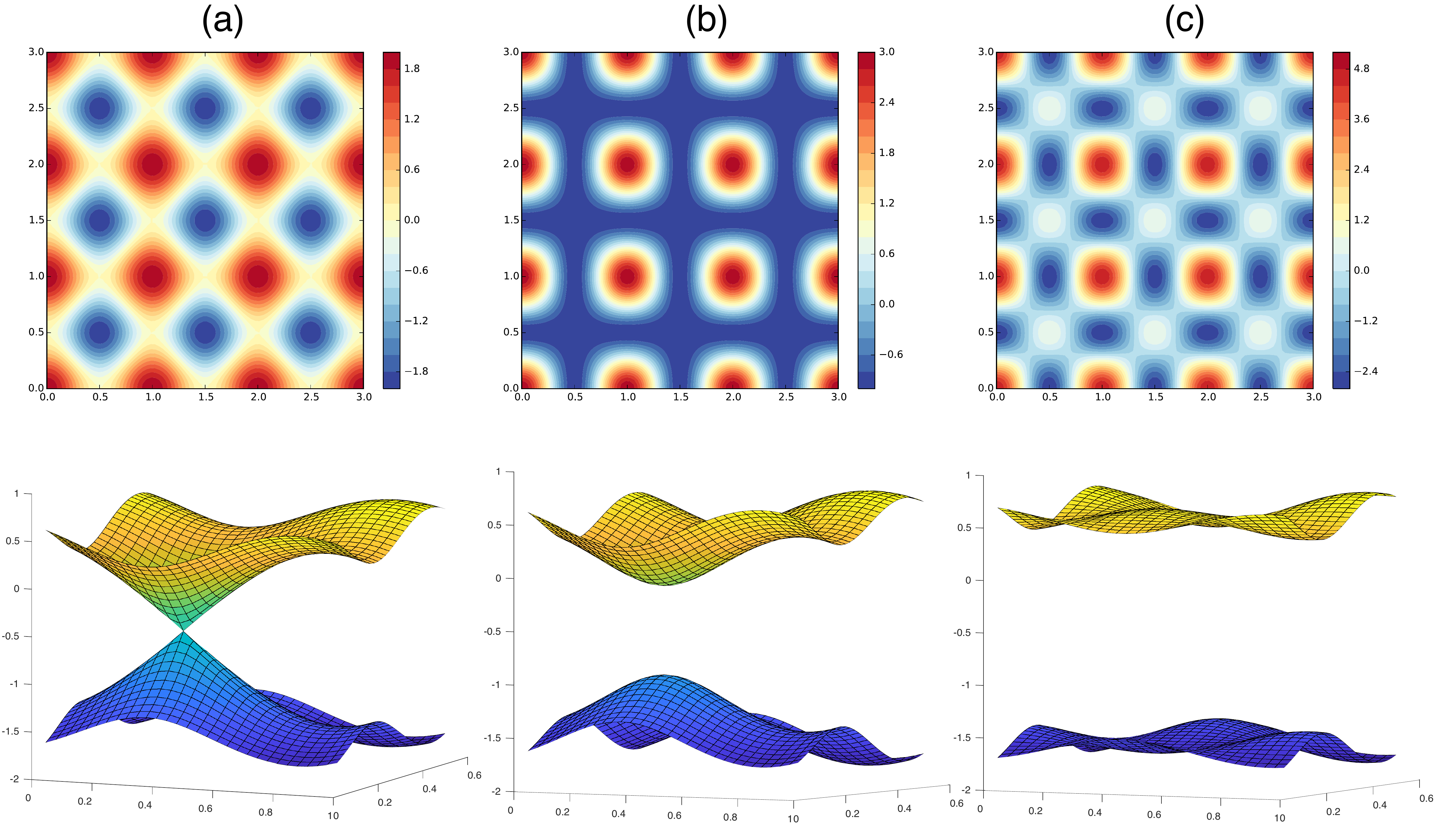}
\caption{\label{Plot_Subband} The lowest Landau level (LLL) under the square lattice potential (Eq.~\eqref{eq:potential}) with the magnetic flux density $\phi=2$ in one unit cell.
We take $V_1=V_2=1$, (a) $V_3=V_4=0$ (b) $V_3=V_4=0.5$, (c) $V_3=V_4=1.4$. The upper panel shows the potential pattern, and the lower panel shows the band structure.
The LLL splits into two sub-bands, the lowest band has Chern number $C=1$ when $V_3=V_4>0$. Tuning the potential pattern gives a nearly flat band as shown in (c).}
\end{figure}

We begin with a concrete example: the transition between a continuum $C = 1/3$ Laughlin state and a $C = 2/3$ FCI.
As before, we  consider a flat continuum Landau level with the addition of square superlattice potential described in Fig.~\ref{Plot_Subband}, but with the addition of a Coulomb interaction.
The transition will be driven by the competition between the lattice amplitude $U_0$ and the Coulomb scale $E_C$.
We consider flux density $\phi = 2$ and  electron density $n = 2/3$, giving filling fraction $\nu = 1/3$. 
At $U_0 = 0$, the electrons fill $1/3$ of the LLL, where interactions stabilize the $\nu = 1/3$ Laughlin state ($C = 1/3, s = 0$).
When $U_0 > 0$, the LLL splits into two subbands, and when $U_0$ is sufficiently strong,  the lower band, which carries $C=1$, will be $2/3$ filled, while the upper band, with $C=0$, will be empty.
For appropriate choices of $V_m$, the subbands are very flat (Fig.~\ref{Plot_Subband}), i.e. the bandwidth $W$ of the subbands is  small compared with the gap $\Delta$ between the subbands. 
Following the philosophy of earlier lattice FCI studies (e.g.  Ref.~\cite{Neupert2011,Sheng2011, Regnault2011,Tang2011}), we expect an $\sigma_{xy} = 2/3$ FCI may appear ($C = 2/3, s = -2/3$) for an appropriate ratio of $U_0 / E_C$.
Consequently $U_0 / E_C$ may drive a direct phase transition between a $\sigma_{xy} = 1/3$ FQH state and the $\sigma_{xy} = 2/3$ FCI.

\begin{figure}[t]
\includegraphics[width=0.49\textwidth]{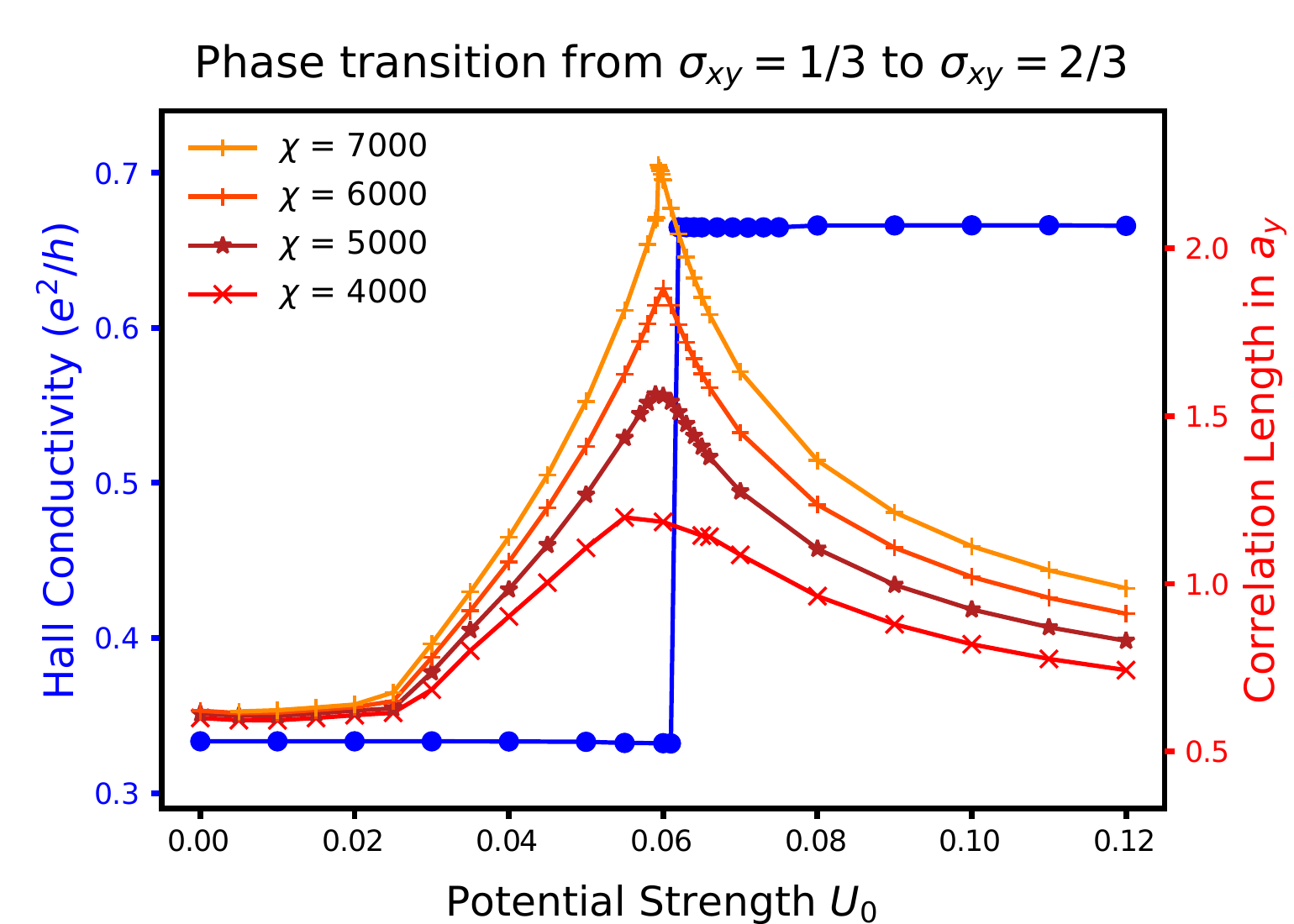}
\caption{\label{Phase_Transition} The DMRG simulation result of the electrons on the infinite cylinder with circumference size $L_y = 31.9 l_B$. 
We consider the LLL under a square lattice potential (Eq.~\eqref{eq:potential}) with $V_1 = V_2 = 1$, $V_3 = V_4 = 1.4$, magnetic flux density (per unit cell) $\phi =2$, electron density $n = 2/3$. 
We change the strength of lattice potential $U_0$ with a step size $\Delta U = 0.001$, 
with strength of the short-ranged coulomb interaction being constant $E_C=1$. 
The blue line represents the change of Hall conductivity, which is measured at bond dimension $\chi=2000$. 
It shows a sharp direct transition between the $\sigma_{xy}=1/3$ and $\sigma_{xy}=2/3$ state at $U_0\sim 0.06$.
Red lines represent the change of correlation length (in the unit of lattice constant $a$ in $y$-direction)  with different bond dimensions $\chi$. 
At the phase transition point ($U_0\sim 0.06$), the correlation seems to diverge with the bond dimensions.
The truncation error is $\epsilon_{\text{trun}} <10^{-6}$ at $U_0=0$, $\epsilon_{\text{trun}} \sim 6\times 10^{-5}$ at the critical point and $\epsilon_{\text{trun}} \sim 4\times 10^{-5}$ at the $\sigma_{xy}=2/3$ phase.
 } 
\end{figure}

\subsection{iDMRG Numerical Simulations}
The existence of continuous transition needs further substantiation since there might very well be a first order transition or an intermediate phase. To this end we use the iDMRG method~\cite{McCulloch2008,White1992} developed for quantum Hall problems in Refs.~\cite{Zaletel2013_FQH, Zaletel2015_multiQH}. 
We wrap the 2D system onto a cylinder which is infinite along $x$-direction but compact along $y$-direction, with circumference $L_y$. 
The Coulomb interaction is taken to have the $k$-space form $V(k) = E_C \frac{2\pi}{k} \tanh(k d)$, where we take into account metallic gates at a distance $d$  which screen the Coulomb interaction, as are present in recent experiments~\cite{FCI_experiment}.
Here we fix $d = 2 a$, twice the lattice spacing. At $\phi = 2$, $a = \sqrt{4 \pi} \ell_B$, so we consider a cylinder of circumference $L_y = 9 a \approx 31.9 l_B$.
The Coulomb interaction strength $E_C$ is fixed to $E_C=1$ while superlattice amplitude $U_0$ is increased from zero in increments of $\Delta U = 0.001$.  
The Hall conductivity is computed by measuring the change of the charge polarization charge during adiabatic flux insertion \cite{Zaletel2014_Flux}. 

In Fig.~\ref{Phase_Transition}, we observe a very sharp phase transition between $\sigma_{xy} = 1/3$ and $\sigma_{xy} = 2/3$ phases driven by $U_0$. The correlation length peaks at the transition and grows with the DMRG bond dimension ``$\chi$.'' While finite-circumference and DMRG accuracy effects make it difficult to conclusively identify whether the phase transition is continuous or  first-order,  several observations are in favor of a continuous transition. 
First, there is no discontinuity in the derivative of energy $\partial E/\partial U_0$ across the phase transition.
Second, unlike the first order metal-FCI transition in Ref.~\cite{Zaletel2015_FCI}, the correlation length appears to diverge on both sides of the transition, and at the transition it increases with the DMRG bond dimension $\chi$ as would be expected from the theory of ``finite entanglement scaling''~\cite{Pollmann2009}.

Though we have studied the square lattice potential most extensively due to its numerical advantages, we note that we observe an analogous  transition for a triangular lattice potential, again at $\phi=2$, $n=2/3$. Therefore this type of transition may be observable in the boron-nitride / graphene heterostructures realized in Ref.~\cite{FCI_experiment}.

To further understand the nature of the phase transition, we will first analyze the possible critical theory using the composite fermion construction, which predicts an emergent SU(3) symmetry. 
In Sec.~\ref{sc:physical_property} we provide numerical evidence for this emergent SU(3) symmetry, which can only arise if the transition is continuous.

\subsection{Composite Fermion construction}
\label{sc:Discussion}
To understand the two phases, the $\sigma_{xy}=1/3$ FQH state and a $\sigma_{xy}=2/3$ FCI, as well as the transition between them, we use the composite fermion (CF) construction.\cite{Jain1989, KolRead1993, Moeller2009,Moeller2015} An extension of this theory to the general case is discussed in Section \ref{sc:familyCFs}. 
The composite fermion is formed by attaching two flux quanta to each electron. The electron and CF Hall conductivities are related by
\begin{equation}
\sigma_{xy} = \sigma_{xy}^{\textrm{CF}}/(1+2\sigma_{xy}^{\textrm{CF}}).
\end{equation}
Thus the $\nu=1/3$ FQH state arises when the CFs form a  $C=\sigma^{\textrm{CF}}_{xy}=1$ integer QH state, while the $\sigma_{xy}=2/3$ state arises when the CFs have Chern number $C=-2$. 
So from the perspective of CFs, the transition is a $C=1$ to $C=-2$ Chern-number changing transition.
This transition arises for the same reason the lattice potential drives such transitions for electrons:
the CFs  also experience a lattice potential, $\mu_{\textrm{CF}}=U_{\textrm{CF}}\int d\bm r \,  \sum_{m=1} ^4  (V^{\textrm{CF}}_me^{ i\bm r \cdot  \bm G_m } n_{\bm r}+h.c.)$, where the amplitude $U_{\textrm{CF}} \propto U_0$ and the amplitudes $V^{CF}_m$ are some effective renormalized constants.
Due to flux attachment, the average flux seen by the composite fermions is $\phicf = \phi - 2 n=2/3$ per unit cell, e.g., they are at filling 1.
Without the potential, the composite fermions fill their LLL (with $C = 1$), generating the $\sigma_{xy}=1/3$ FQH state.
But once the potential strength $U_{\textrm{CF}}$ is comparable to the composite fermion cyclotron gap $\omega_{\textrm{CF}}$, the CF-bands can touch and the the Chern number may change to $-2$, giving the $\sigma_{xy}=2/3$ FCI state, similar to the free fermion example  discussed in Fig.~\ref{fig:3Dirac}. Such a $\Delta C = 3$ transition relies on the fact that the CFs see flux $\phicf = 2/3$ even while the electron flux is $\phi = 2$.

At the transition point, the composite fermions will form three Dirac cones which are protected by the magnetic translation symmetry (since $\phicf=2/3$).
The critical theory is 
\begin{align}\label{eq:FQH_FCI}
\mathcal L &= \sum_{I=1}^{3}  \bar \psi_I  (i\slashed{\partial} + \slashed{a}) \psi_I -\frac{1}{8\pi} a d a + \frac{1}{8\pi} (a-A) d (a-A) \nonumber \\
& = \sum_{I=1}^{3}  \bar \psi_I   (i\slashed{\partial} + \slashed{a}) \psi_I -\frac{1}{4\pi} A da +\frac{1}{8\pi} A dA.
\end{align}
Here, $\psi_I$ are the two-component Dirac fermions interacting with a dynamical $U(1)$ gauge field $a$ which arises from  flux attachment. A Maxwell-like term for the gauge field $a_\mu$ is also present, which we suppress in the formal Lagrangian for notational convenience.
$A$ is the probe (external) gauge field which couples to the charge of the electrons. For notational simplicity, we implicitly measure $A$ and the electron density relative to their  values at the transition $A_\ast, n_\ast$; this could be made explicit as in Eq.~\eqref{eq:free_dirac}, 
to account for the change in ``$s$,'' but these terms have no influence on the dynamics.
The first two terms are very similar to the free fermion case in Eq.~\eqref{eq:free_dirac} (with $\bar{C} = -\frac{1}{2}$), the only difference being that the Dirac fermions are interacting with the dynamical $U(1)$ gauge field instead of a static background field.
Meanwhile, the third  Chern-Simons term encodes the flux attachment constraint. 

The transition theory is a pure QED$_3$ with three flavors of Dirac fermions. and the tuning transition parameter is the mass term of composite fermions, $m \sum_I  \bar \psi_I \psi_I$.
Once the composite fermions are gapped,
integrating them out gives an effective theory depending on the sign of $m$:
\begin{eqnarray}
{\cal L}' &=& \frac{m}{|m|} \frac{3}{8\pi} a da -\frac{1}{4\pi} A da +\frac{1}{8\pi} A dA.
\label{eq:charge}
\end{eqnarray}
After integrating out $a$, we obtain  ${\cal L}_{\textrm{eff}} = \frac{3 - \text{sign}(m)}{6} \frac{1}{4\pi} AdA$. Therefore the theory describes the $\sigma_{xy} = 1/3$ phase for $m>0$ and $\sigma_{xy} = 2/3$ phase for $m<0$.

Before closing this section we mention a technical aside. 
Strictly speaking, we should write Eq.~\eqref{eq:FQH_FCI} a little differently for it to be free of the parity anomaly. First note, the flux attachment proceedure can be alternatively formulated using a parton construction, writing the electron as $c = b \psi$, here $b$ is a gauge charged (and electrically charged) boson. Putting the boson $b$ in the $\nu=1/2$ Laughlin state is equivalent to attaching two flux quanta to the electron $c$, and the fermionic parton $\psi$ is then nothing but the composite fermion.
If we represent the three current of  $b$ as $ d\wedge \beta/2\pi$, then we obtain:
\begin{equation}
\mathcal L = {\mathcal L}_{\textrm{CF}}[a,\psi] +\frac{1}{2\pi} (a-A) d \beta - \frac{2}{4\pi} \beta d \beta
\end{equation}
Now we put the composite fermions $\psi$ into a Dirac band structure such as Fig.~\ref{fig:3Dirac}, we have
\begin{equation}
\mathcal L_{\textrm{CF}}[a, \psi] = \sum_{I=1}^{3}  \bar \psi_I  (i\slashed{\partial} + \slashed{a}) \psi_I -\frac{1}{8\pi} a d a
\end{equation}
The theory is now free of parity anomaly, and the Chern-Simons terms are properly quantized.
Finally `integrating out' $\beta$ will lead to Eq.~\eqref{eq:FQH_FCI}.
The last step requires some care, but those finer points do not affect the dynamics of the transition.

\section{A family of QED3-Chern-Simons theory with arbitrary Dirac flavor}\label{sc:family}

\subsection{Composite fermion construction for the transition}
\label{sc:familyCFs}
The analysis above can be extended to a more general framework for  phase transitions between FCI states.
In general the CF is obtained by attaching $k$ flux quanta to the electrons, with $k$ odd for bosons and $k$ even for fermions.
After flux attachment, the CFs will see an effective flux density $\phi_{\textrm{CF}}=\phi- n k$, and still have particle density $n$.
FCIs arise when the composite fermions form integer Chern-insulators satifying  $n = C_{\textrm{CF}} \phicf + s_{\textrm{CF}}$ for $C_{\textrm{CF}}, s_{\textrm{CF}}$ integers.\cite{KolRead1993, Moeller2009} From these relations the fractional $C, s$ invariants of the electrons are  related to those of the CFs by
\begin{align}
\sigma_{xy}=C = \frac{C_{\textrm{CF}}}{ k C_{\textrm{CF}} + 1 }, \quad s = \frac{s_{\textrm{CF}}}{k C_{\textrm{CF}} + 1 }.
\end{align}
The phase transition between two states in the same Jain sequence, with $\sigma_{xy}=C_1/(kC_1+1)$ and $\sigma_{xy}=C_2/(kC_2+1)$, is then understood as a CF Chern-number  changing transition from $C_1$ to $C_2$.
As we discussed before, the Chern-number changing transition will be accompanied by a gap closing at which $|C_2-C_1|$ Dirac cones  emerge. 
The resulting effective theory for a transition  is a straightforward generalization of Eq.~\eqref{eq:FQH_FCI}:
\begin{align}\label{eq:critical}
{\mathcal L} &= \sum_{I=1}^{|C_2-C_1|}  \bar \psi_I   (i\slashed{\partial} + \slashed{a} - m ) \psi_I + \frac{C_2+C_1}{8\pi} a d a 
\nonumber \\&
+ \frac{1}{4 k \pi} (a-A) d (a-A) .
\end{align}

In general we have $N_f=|C_2-C_1|$ Dirac fermions interacting with a Chern-Simon term at the level $K=(C_2+C_1)/2+1/k$.
The Chern-Simons term  in Eq.~\eqref{eq:critical} is not properly quantized when $k\neq 1$. 
As for $N_f=1$ QED$_3$, this can be fixed by declaring that only the monopoles which create multiples of $2 k \pi$ flux are allowed in the theory~\cite{wangsenthil15b,MaxAshvin15}:
in a bosonic theory (odd $k$) the minimal monopole is a local boson which carries integer Lorentz spin, while in a fermionic theory (even $k$), the minimal monopole is a fermion with half integer spin.
Alternatively, we may introduce an auxiliary Chern-Simons field ($b_\mu$) and rewrite $\frac{1}{4k\pi} (a-A) d (a-A)$ as $\frac{1}{2\pi} b d(a-A) - \frac{k}{4\pi} b d b$~\cite{Seiberg2016395}.
This is similar to the parton formulation of  flux attachment  discussed in previous section.
Or finally, we may redefine the gauge charge of the Dirac fermions, $a \rightarrow k a$~\cite{SeibergLargeCharge}.

The mass term $m$ is again the tuning parameter for the phase transition.
When $m\gg 1$, integrating out the $\psi_I, a$ produces an FCI with $\sigma_{xy}=C_2/(kC_2+1)$, while for $m\ll -1$,  $\sigma_{xy}=C_1/(kC_1+1)$.
Tuning $m$ to a critical value ($m_c \sim 0$) results in a critical point between two FCI/FQH phases, which we dub  QED$_3$-Chern-Simons theory.
When $K = 0$,  the critical $m_c$ would be pinned to exactly $0$ by the emergent time-reversal symmetry (microscopically it is an emergent particle-hole symmetry).
For $K \neq 0$, the critical mass $m_c$ is  finite and will depend on microscopic details.

As we discussed before, a direct transition requires that the other mass terms $\bar \psi_i M_{ij} \psi_j$ ($M\neq$ Identity) are forbidden by the magnetic translation symmetry arising from the fractional $\phicf = p / (C_2 - C_1)$ (with $p$ and $C_2-C_1$ being co-prime).
In terms of the electron density $n$ and  flux density $\phi$, this constraint becomes
\begin{eqnarray}
n& = & p C_2 / (C_2-C_1)  \mod 1,  \\ 
\phi & = & k\cdot n+\phicf = p (k C_2+1)/(C_2-C_1) \mod 1. \quad \quad
\end{eqnarray}
Thus, to recover for example the $\sigma_{xy}=1/3$ to $2/3$ transition we set $C_1=1$ and $C_2=-2$, which gives $n = 2p/3$ and $\phi = p$ (modulo integers). This is consistent with the situation described in Section \ref{sc:Model}, where $n=2/3$ and $\phi=2$.

The other potentially relevant operator is the monopole operator.
Due to the mutual Chern-Simons term $adA$, the monopole operator carries  non-zero electrical charge.
Therefore monopole operators are forbidden due to charge conservation, leaving the critical point intact.

\subsection{Pure QED$_3$ theory with $N_f$ flavors}
We call theories without a Chern-Simons term for $a$ ($K = 0$) ``pure'' QED$_3$. The family of critical theories in Eq.~\eqref{eq:critical} contains pure QED$_3$ with any number $N_f$ of Dirac fermions, with $N_f$ both even and odd.

Pure QED$_3$ with odd-$N_f = 2 N + 1$  is realized when $k=2$, $C_1=-N-1$, $C_2=N$.
Therefore it can appear in a system made of fermions, with  critical theory 
\begin{equation}\label{eq:fermionic_QED3}
 \mathcal L= \sum_{I=1}^{2 N+1}  \bar \psi_I   (i\slashed{\partial} + \slashed{a}) \psi_I -\frac{1}{4\pi} A da +\frac{1}{8\pi} A dA .
\end{equation}
This describes the transition between the $\sigma_{xy}=N/(2N+1)$ FCI and $\sigma_{xy}= (N+1)/(2N+1)$ FCI, which are the particle-hole symmetric partners in the fermionic Jain sequence.
The required CF  flux is $\phicf=p/(2N+1)$, where $p$ is co-prime with $2 N + 1$.
Meanwhile, the fermion density is  $n \equiv C_2 \cdot\phicf \equiv p N / (2N+1) \mod 1$. Thus, the magnetic flux in the original fermionic system is $\phi= 2\cdot n + \phicf = p  \mod 1$, 
which is an integer.

When $N=0$, the transition is between a $\sigma_{xy}=0$ trivial state and a $\sigma_{xy}=1$ CI. 
Here Eq.~\eqref{eq:fermionic_QED3}  reduces to the vortex dual of a single free Dirac fermion~\cite{sonphcfl,wangsenthil15b,MaxAshvin15}.
 $N=1$ gives the $N_f=3$ QED$_3$ discussed in Sec~\ref{sc:FQH_FCI}.

Pure QED$_3$ with $N_f$-even is realized when $k=1$, $C_1=-N-1$, $C_2=N-1$.
Therefore it can appear in a bosonic system, with critical theory
\begin{equation}\label{eq:bosonic_QED3}
 \mathcal L= \sum_{I=1}^{2N}  \bar \psi_I   (i\slashed{\partial} + \slashed{a}) \psi_I -\frac{1}{2\pi} A da +\frac{1}{4\pi} A dA  .
\end{equation}
This describes the transition between a $\sigma_{xy}=(N-1)/N$ FCI and a $\sigma_{xy}= (N+1)/N$ FCI, which are particle-hole symmetric partners in the bosonic Jain sequence~\cite{WangSenthil16,Geraedts17}.
To protect the transition the CFs must see flux $\phicf= p /(2N)$, where $p$ is co-prime to $2N$. Meanwhile, the boson density should be  $n\equiv C_2\cdot\phicf \equiv  p (N-1)/(2N) \mod 1$. Thus, the magnetic flux in the original bosonic system is $\phi= k\cdot n + \phicf =1/2 \mod 1$.

When $N=1$, this theory reduces to the self-dual   $N_f=2$ QED$_3$ that describes the transition between the $\sigma_{xy}=0$ (Mott insulating phase) and the $\sigma_{xy}=2$ bosonic integer quantum Hall state~\cite{Tarun2013_QHtransition,YML2014_SPT}. 
This transition can be realized at  $\phi=1/2$ and  $n=1$.
Another interesting case is $N=2$,  which describes the transition between the $\sigma_{xy}=1/2$ FCI and the $\sigma_{xy}=3/2$ FCI.
$N_f=4$ QED$_3$ also arises as the effective theory of a Dirac spin liquid \cite{Hastings2000,HermeleKagome}) . 
This transition arises at $\phi=1/2$ and $n=1/4$.
Such a transition may be possible to realize in a cold-atomic system (e.g. bosonic Harper-Hofstadter model~\cite{Moeller2009,YCH_hofstadter}).

\section{Physical properties}\label{sc:physical_property}

\subsection{Critical exponents in large-$N_f$ limit}

We now examine some of the physical properties of the resulting critical theory which may be physically measurable. These properties depend on flavor number of Dirac fermions $N_f$ and the total Chern-Simons coefficient $K$ for  $a$.

It is interesting to consider some limits of the critical theory Eq.~\eqref{eq:critical} in which one can hope for some quantitative understanding. One familiar limit is the large-$N_f$ limit, in which $N_f \to \infty$ while the ratio $\lambda\equiv 8K / \pi N_f$ is held fixed. In this limit, one can calculate many critical exponents and transport properties in a controlled manner.
Usually large-$N_f$  is considered an artificial limit not directly related to any physical system. 
Here, however, theories with large $N_f$ correspond to a sequence of quantum critical points that in principle can be realized in experiments. This opens the possibility of comparing well established large-$N_f$ theoretical calculations with future experimental observations.

The scaling dimensions of fermion mass operators, to leading nontrivial order in $1/N_f$, 
can be calculated for arbitrary $\lambda$~\cite{Chen1993_transition,RantnerWen2002,Hermele2005,Hermele2005_Erratum,Hui2018} 
We find that the scaling dimension of the SU$(N_f)$ adjoint mass operators in Eq.~\eqref{adjointmass}, which correspond microscopically to the CDW order parameters with momenta given by Eq.~\eqref{momenta}, is given by Eq.~\eqref{momenta}, have scaling dimension
\be \label{eq:Scaling_adjoint}
\Delta_M=2-\frac{64}{3\pi^2(1+\lambda^2)N_f}+O(1/N_f^2).
\ee
The SU$(N_f)$ singlet mass $\sum_i\bar{\psi_i}\psi_i$, which is the operator one tunes across the transition, has dimension
\be \label{eq:Scaling_singlet}
\Delta_m=2+\frac{128(1-2\lambda^2)}{3\pi^2(1+\lambda^2)^2N_f}+O(1/N_f^2) ,
\ee
which corresponds to
\be
\nu=1+\frac{128(1-2\lambda^2)}{3\pi^2(1+\lambda^2)^2N_f}+O(1/N_f^2) .
\ee
Some details of the calculation can be found in Appendix~\ref{Appendix:LargeN}. Notice that at $O(1/N_f)$, for $\lambda<1$ the adjoint mass is more relevant, while for $\lambda>1$ the singlet mass becomes more relevant. 
Of course, translation symmetry prevents the appearance of the adjoint mass terms in the Lagrangian. 

The scaling dimension of monopoles has been calculated at the large-$N_f$ limit for both $\lambda=0$~\cite{kapustinqed,DyerMonopoleTaxonomy} and $\lambda\neq 0$~\cite{ChesterPufuChern-Simons}.
In our critical theory, the minimal monopole operator corresponds to the single electron or boson operator.
Therefore, their scaling dimension may be detected using  STM spectroscopy, as will be discussed in more detail in  Sec.~\ref{sc:monopole}.

\subsection{Conductivity tensor \label{sc:conductivity}}

We  consider the conductivity tensors $\sigma_{ij}$ at the critical points, which are expected to take some universal values.  
It is easy to calculate $\sigma_{ij}$ if $N_f$ is large, in which case the gauge field fluctuation is suppressed and an RPA-type calculation (familiar in the context of composite fermi liquid\,\cite{HLR1993}) is enough to determine  the results to sub-leading order in $1/N_f$. 
The electric conductivity tensor is given by
\be
\sigma=\frac{1}{k^2}\left(\sigma_{Dirac}+K\hat{\epsilon} \right)^{-1}+\frac{1}{k}\hat{\epsilon},
\ee
where $\sigma_{Dirac}$ is the conductivity tensor of the Dirac fermions, and $\hat{\epsilon}$ is the antisymmetric tensor with $\epsilon_{xy}=1$ and  recall that $k$ is the number of statistical flux quanta attached to the composite fermion.
If we neglect gauge fluctuation beyond the RPA order (justified if $N_f$ is sufficiently large), then we can replace $\sigma_{Dirac}$ by its free-fermion value $\sigma_{FD}$. 
The full conductivity (in unit of $e^2/h$) tensor will then be
\begin{eqnarray}
\sigma_{xx}&=&\frac{1}{k^2}\frac{N_f\sigma_{FD}}{\left(N_f\sigma_{FD}\right)^2+K^2}, \nonumber \\
\sigma_{xy}&=&\frac{1}{k}-\frac{1}{k^2}\frac{K}{\left(N_f\sigma_{FD}\right)^2+K^2}.
\end{eqnarray}
The actual value of $\sigma_{FD}$, in the scaling regime with short-range disorder, is a universal function of the two ratios $\omega/\eta$ and $T/\eta$, where $\eta$ is the effective elastic scattering rate of the Dirac fermions\,\cite{RyuLudwig07}. In the optical limit $\omega\gg T\gg\eta$, $\sigma_{FD}=\pi/8$. In the DC limit ($\omega=0$) at low temperature $T\ll\eta$, $\sigma_{FD}=1/\pi$ -- this result, however, should be taken with caution since at low temperature, disorder may become relevant and eventually drive the critical point away from the clean limit (see, for example, Ref.~\cite{ThomsonSachdev2017}).

Notice that if $K=0$, then $\sigma_{xy}=1/k$ is an exact result due to an emergent particle-hole (or time-reversal) symmetry of the critical theory.
 Finally, notice that at the free fermion transition, 
\begin{equation}
\sigma=\Delta C\sigma_{FD}+\bar{C}\hat{\epsilon},
\end{equation}
where $\Delta C$ is the change of total Chern number and $\bar{C}$ is the average of Chern numbers of the two nearby states.

\subsection{Emergent SU$(N_f)$ symmetry and charge-density waves \label{sc:CDW}}

The Lagrangian in Eq.~\eqref{eq:critical} apparently has an emergent SU$(N_f)$ flavor symmetry which rotates the Dirac fermions:
\be
\psi_I\to U_{IJ}\psi_J,
\ee
where $U\in \SU(N_f)$. Such a symmetry is absent microscopically -- the magnetic translation symmetry acts on the continuum fermion fields as a subgroup of the $\SU(N_f)$ flavor group (Eq.\eqref{mtrans}). Therefore terms that violate the $\SU(N_f)$ symmetries (but preserve the magnetic translation symmetry) are allowed. Such terms at lowest order contain four-fermion fields. The emergence of the flavor $\SU(N_f)$ symmetry at the fixed point requires the irrelevance of such four-fermion terms, which is true if $N_f$ is sufficiently large\,\cite{Hermele2005}. The exact value of a critical $N_f$ (as a function of $K$) is not known. In the following, we will assume that $N_f$ is not too small and the $\SU(N_f)$ symmetry does emerge in the deep infra-red, and obtain consequences of this enlarged symmetry.

Following the spirit of Ref.~\cite{Hermele2005}, the emergent $\SU(N_f)$ symmetry relates many operators with very different microscopic origins. The simplest such gauge-invariant operators are the fermion mass bilinears that form an $\SU(N_f)$ adjoint representation:
\be
\label{adjointmass}
\bar{\psi}_IM_{IJ}\psi_J,
\ee
where $M$ is an $N_f\times N_f$ invertible traceless Hermitian matrix. There are in total $N_f^2-1$ independent mass bilinears in the $\SU(N_f)$ adjoint representation. 
The emergent $\SU(N_f)$ symmetry implies they all have the same scaling dimensions $\Delta_{M}$. 
What do these operators correspond to microscopically? Assuming the CFs transform under translation as in Eq.~\eqref{mtrans}~\footnote{Strictly speaking each symmetry operation on the CFs is ambiguous up to an overall phase, because they couple to a U(1) gauge field, but this ambiguity cancels for the bilinears.}, it is straightforward to see that these operators transform nontrivially under translation symmetry, since the Dirac cones are at different momenta.
Notice that since the mass bilinear operators do not carry a physical electric charge, they do not see any magnetic flux. Thus, they will transform under the usual translation symmetry ($\mathbb{Z} \times \mathbb{Z}$) rather than the magnetic translation symmetry. 
Following Eq.~\eqref{mtrans} we find that the bilinears  carry lattice momenta
\begin{eqnarray}
\label{momenta}
(k_x, k_y)&=&\left(\frac{2\pi n_x}{N_f}, \frac{2\pi n_y}{N_f} \right), \nonumber \\
 n_x, n_y&\in& \{0, 1, ... N_f-1\}, \hspace{5pt} (k_x,k_y)\neq(0,0),
\end{eqnarray}
and there are exactly $N_f^2-1$ different momenta of the CDW order parameter, matching the number of independent $\SU(N_f)$ adjoint mass operators. We therefore interpret the $\SU(N_f)$ adjoint mass operators as charge-density wave (CDW) order parameters $\rho_{(k_x,k_y)}$ at different momenta. Alternatively,  notice that the mass operators preserve translations by $N_f$ unit lengths in either the $\hat{x}$ or $\hat{y}$ direction. So these operators transform under the quotient group $(\mathbb{Z}\times\mathbb{Z})/(N_f\mathbb{Z}\times N_f\mathbb{Z})=\mathbb{Z}_{N_f}\times\mathbb{Z}_{N_f}$, for which  there are exactly $N_f^2-1$ different nontrivial representations.
The mass matrix for the CDW order parameter indexed by $n_x$, $n_y$ is $M_{IJ}=\delta_{I-J,n_x} \exp(i n_y 2\pi I/N)$ in the gauge used in Eq.~\eqref{mtrans}.

In principle, one can turn on an additional symmetry-breaking periodic potential $V_{\vec{k}}$ at the special momenta of Eq.~\eqref{momenta} and measure the resulting response in the CDW order parameter $\rho_{\vec{k}}$. The emergent $\SU(N_f)$ symmetry implies that near the critical point the universal part of the response will be identical for all of the $N_f^2-1$ momenta in Eq.~\eqref{momenta}. More specifically, near the transition we expect a linear response $\rho_{\vec{k}}=\chi_{\vec{k}}V_{\vec{k}}$ where the susceptibility $\chi_{\vec{k}}$ diverges as $\chi_{\vec{k}} \sim |m-m_c|^{-\gamma_{\vec{k}}}$ as one approaches the critical point at $m_c$. From the emergent $\SU(N_f)$ symmetry, the exponents $\gamma = \gamma_{\vec{k}}$ of the special momenta are identical. Exactly at the critical point, we have $\rho_{\vec{k}} \sim (V_{\vec{k}})^{1/\delta_{\vec{k}}}$, where the $\delta = \delta_{\vec{k}}$ of the special momenta are identical. One can also measure the response slightly away from the special momenta: $\vec{k}=\frac{2\pi}{N_f}(n_x,n_y)+ \Delta \vec{k}$. Then one expects a linear response with $\chi_{\vec{k}} \sim| \Delta \vec{k} |^{2 \Delta_M-3}$, again with identical exponents for all $(n_x,n_y)$, where $\Delta_M$ is the scaling dimension of the adjoint mass operator. The standard scaling relations among these exponents are $\delta=(3-\Delta_M)/\Delta_M$ and $\gamma=\nu(3-2\Delta_M)$, where $\nu$ is the correlation length exponent. If $\Delta_M>3/2$ (as is the case for large $N_f$), one should include a smooth non-universal piece into the susceptibility which may numerically dominate over the universal singular piece. In this case the smooth part of the response must be subtracted to probe the universal physics.

\subsubsection{Numerical evidence for the emergent symmetry}
We return to the iDMRG results of Sec.~\ref{sc:FQH_FCI} to look for evidence of an emergent $\SU(N_f)$ symmetry.
Since $N_f = 3$, the critical theory will have 8 adjoint mass terms which can be probed by the Fourier-transformed electron density $\rho_{\vec{k}}$ for $\vec{k} \in \Lambda$, where $\Lambda = \{ (\frac{2 \pi n_x }{3}, \frac{2 \pi n_y}{3}) \}$ are the special momenta.
To isolate these operators, we suppose the real space density $\rho(r)$ has a decomposition of the form $\rho(r) = \sum_{\vec{k} \in \Lambda} e^{i \vec{k} \cdot \vec{r}} \rho_{\vec{k}} (r) + \cdots$, where the $\rho_{\vec{k}} (r)$ vary slowly over the lattice scale and the neglected terms decay more rapidly.
In the 2D limit, the emergent $\SU(3)$ symmetry can then be probed by the equal-time correlation function $C_{\vec{k}}(r) =  \langle \rho_{-\vec{k}} (r) \rho_{\vec{k}} (0) \rangle$.
At the putative critical point, $C_{\vec{k}}$ should show a power-law decay $C_{\vec{k}}(r) \propto r^{-2 \Delta_{\vec{k}}}$.
The non-trivial prediction of $\SU(3)$ symmetry is that the scaling dimensions $\Delta_{\vec{k}}$ of the 8 distinct $\vec{k} \neq 0$ momenta are identical.

\begin{figure}[p]
\centering
\includegraphics[width=0.35\textwidth]{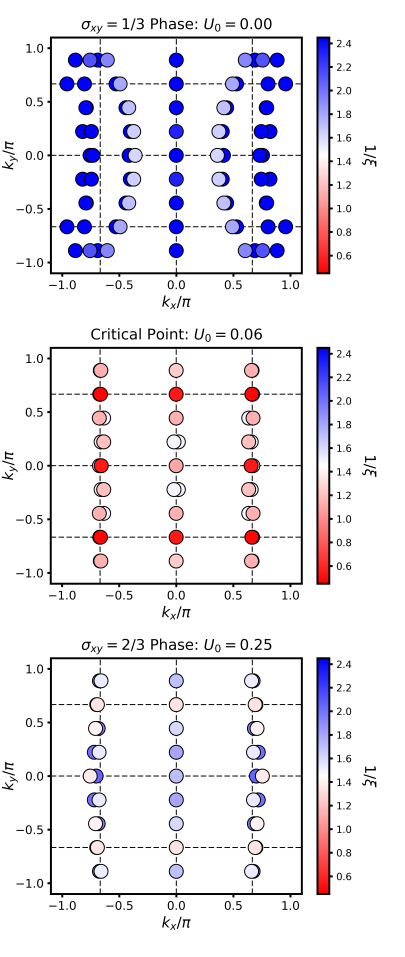} 
\vspace{-10pt}

\caption{\label{TMspectrum} Correlation length spectra of operators carrying zero electric charge and crystal momenta $\vec{k}$ in three different regimes: (i) $\sigma_{xy} = 1/3$ FQH phase (ii) Critical point (iii) $\sigma_{xy} = 2/3$ FCI phase. The spectrum is obtained from simulations with DMRG bond dimension $\chi=6000$. The color bar represents the inverse of correlation length $1/\xi$. Note that we have a finite circumference size $L_y=9$ in the units of the lattice constant, so the $k_y$ take discretized values $k_y = 2\pi n/L_y$, $n = 0,1,\dots L_y -1$.
At the critical point, the correlation lengths at the eight momenta $(k_x, k_y)=(2\pi n_x/3, 2\pi n_y/3) \neq 0$ are the largest and identical. These modes correspond to the eight $\SU(3)$ adjoint mass terms. 
Numerically, correlation lengths at other $k$-points have at least twice smaller correlation lengths at $\chi=7000$, and increase much slower under entanglement scaling compared to correlation lengths at these eight momenta.
The spectra at the $\sigma_{xy}=1/3$ and $2/3$ phases have no additional symmetries other than two reflections ($k_x$ and $k_y$) inherent to the simulation. In the case of $\sigma_{xy} = 2/3$, one may notice there are also eight points with almost identical correlation lengths near aforementioned eight special momenta. However, $i$) they deviate from the exact momenta of $\SU(3)$ adjoint mass terms, and $ii$) their correlation lengths are not well separated from correlation lengths of other $k$-points, e.g. $(k_x,k_y) \sim (\pm 2\pi/3, \pm 4\pi/9)$. 
}
\end{figure}

	However, the interpretation of our numerics is complicated  both by the finite circumference of the cylinder and the finite accuracy of the DMRG (as characterized by the bond dimension)\cite{MPS_review}, introducing a length scale which cuts off the correlations.
At long distances along the cylinder, the correlation functions will instead decay exponentially with correlation lengths $\xi_{\vec{k}}$.
Luckily we can also use these correlation lengths to probe the emergent $\SU(3)$ symmetry.
If the DMRG is sufficiently accurate, the only length scale at the critical point is the circumference of the cylinder $L$, and hence conformal invariance requires $\xi_{\vec{k}} =  L / \alpha_{\vec{k}} + \mathcal{O}(L^0)$ for some coefficients $\alpha_{\vec{k}}$.
Using the putative conformal invariance, we can exchange the role the infinite spatial direction and time ($x \leftrightarrow \tau$) to reinterpret the equal-time correlation functions of the cylinder as a two point function in imaginary time, $C_{\vec{k}}(\tau = 0, x, y) = C_{\vec{k}}(x / v, 0, y)$, where $v$ is a velocity. 
In this alternative view, $v \xi_{\vec{k}}^{-1}$ is the \emph{energy} of the lowest-energy excited state carrying the corresponding quantum number~\footnote{On the $S^2\times R$ geometry, the state-operator correspondence would imply that $\xi_{\vec{k}}^{-1}$ is the scaling dimension of the most relevant primary operator carrying quantum number $\vec{k}$, up to a normalization factor.  But here we are simulating a cylinder/torus geometry instead of the sphere, so there is no precise correspondence of this form. Nevertheless the \emph{equality} of the $\xi_{\vec{k}}$ will follow from the emergent $\SU(N_f)$}.
If there is an emergent $\SU(3)$ symmetry, these states must transform under the adjoint representation of $\SU(3)$, and hence we should find the ``energies'' $\xi_{\vec{k}}$ are identical.

One can readily obtain the correlation lengths $\xi_{\vec{k}}$ along the cylinder using the DMRG ``transfer matrix technique'' ~\cite{MPS_review, Zauner2015}.
Fig.~\ref{TMspectrum} shows the dominant correlation lengths $\xi$ of charge-neutral operators that carry the crystal momenta $\vec{k}$.
At the critical point, the eight $\vec{k} \neq 0 \in \Lambda$ are found to have nearly the same correlation length.
Microscopically, there is no symmetry reason for these $\xi_{\vec{k}}$ to be similar (except as four pairs related by the inversion symmetry).
These eight points have the smallest $\xi^{-1} \sim 0.5$ (largest $\xi$), with the next smallest value $1/\xi\sim 1.1$  located at $\vec{k} = (0,0)$, which presumably corresponds to the $\SU(3)$ singlet mass term. 
We take the near-equality between the eight $\xi_{\vec{k}\neq 0}$ which are nevertheless distinct from $\xi_{\vec{k} = 0}$ as  evidence for the emergence of $\SU(3)$ symmetry.
Moreover, the large-$N_f$ calculation of the scaling dimensions in Eq.~\eqref{eq:Scaling_adjoint},\eqref{eq:Scaling_singlet} agrees with our observation that the $\SU(3)$ singlet mass term decays faster (smaller $\xi$) than the $\SU(3)$ adjoint mass terms as $\Delta_m > \Delta_M$. 
Once we tune away from the critical point (Fig.~\ref{TMspectrum}), the correlation length becomes  smaller and there is no emergent $\SU(3)$ symmetry.

\subsection{Magnetic monopoles} \label{sc:monopole}

In compact QED$_3$, magnetic monopoles which insert $2\pi$ units of flux of the emergent gauge field represent an fundamental class of local operators which,  unlike other gauge invariant operators such as fermion bilinears, cannot be written as polynomials of the fields. Nevertheless  they correspond to {\em local} operators.  
While we have previously shown that  fermion bilinears  $\bar \psi M  \psi$ correspond to  CDW order at different wave vectors, in this section we identify the physical operators corresponding to  magnetic monopoles. In fact, for the theory that describes the $\sigma_{xy}=1/3$ to $\sigma_{xy}=2/3$ transition we argue that they  correspond to the electron creation/destruction operator, which can be measured by experimental probes like Scanning Tunneling Microscopy (STM).

Consider a general strength-$q$ monopole which creates $2q\pi$ flux of the $U(1)$ gauge field $a_\mu$, in a  theory with $N_f$ massless Dirac fermions and Chern-Simons level $K$. To avoid the parity anomaly in conventional QED3-Chern Simons theories, $N_f/2 + K$ must be an integer~\cite{kapustinqed}. However, as discussed at the end of Section \ref{sc:Discussion}, we will discuss seemingly anomalous theories where $N_f/2 + K=1/k, \mod 1$, which are consistent as long as the allowed monopoles have strength that in multiples of $k$. A particularly simple picture of monopole operators arise in the large $N_f$ limit in the radial quantization picture, where one considers Dirac fermions on the surface of a sphere pierced by $2\pi q$ units of magnetic flux  ~\cite{kapustinqed,DyerMonopoleTaxonomy,ChesterPufuChern-Simons}. The magnetic flux gives rise to $q$ zero modes for each fermion species. If $K=0$, gauge invariance demands that we fill half of these zero modes. Below we will specialize to the case of the $\sigma_{xy}=1/3$ to $\sigma_{xy}=2/3$ transition, for which $N_f=3$, $K=0$ and the allowed monopoles have strength that are multiples of $k=2$. The generalization to other transitions will be discussed elsewhere.

First, let us determine the spin, statistics and global symmetries transformation properties of the monopole operators with strength $k=2$. Note, this leads to two zero modes (labeled $m=\pm 1$), which transform as spinors under rotation,  for each of the $N_f=3$ flavors ($a=1,\,2,\,3$). Filling half of these six zero modes implies there are $_6 C_3=20$ monopole operators which can be written as $\psi^\dagger_{m_1a_1}\psi^\dagger_{m_2a_2}\psi^\dagger_{m_3a_3}|{\rm vacc}\rangle$. Importantly, these are {\em fermionic} operators as can be seen from their half integer spin, and as discussed in \cite{MaxAshvin15,wangsenthil15b}. Furthermore they carry unit charge under the global $U(1)$, as can be seen from the mutual Chern-Simons term in Eqn.~\eqref{eq:charge}. 
Clearly, the only physical local operator that is a charged fermion is the electron. Finally let us discuss the transformation of the monopoles under the $\SU(N_f=3)$ flavor symmetry and Lorentz spin of the 20 strength$-2$ monopoles which are determined from the pattern of filling of the zero modes.
After some algebra, the monopoles can be grouped into two categories: i) Adjoint of SU$(3)$, Lorentz spin $S=1/2$; ii) Singlet of SU$(3)$, Lorentz spin $S=3/2$. As a consistency check, note that the SU$(3)$ adjoint rep has dimension $D=8$, and the singlet of SU$(3)$ has dimension $D=1$. 
Combining this with the $2S+1$ spin degeneracies, we recover $2\times8+4=20$ monopoles in total. 
In the large-$N_f$ limit~\cite{DyerMonopoleTaxonomy,kapustinqed}, the strength-2 monopole has scaling dimension $0.673N_f$.
By naively taking $N_f=3$, we get a scaling dimension $\Delta\approx 2.019$.
The degeneracy between the 16 adjoint and the 4 singlet monopoles will split with higher order correction included. (We note that the $1/N_f$ correction obtained in Ref.~\cite{DyerMonopoleTaxonomy} may not apply here.)

The existence of $16 + 4$ electron-like monopole operators has some interesting experimental consequences, so we analyze their relation to the electron in further detail.
The local electron operator $c$ should have overlap with all symmetry-allowed local operators of the CFT, leading to a decomposition of the general form
\begin{align}
\label{eq:cdecomp}
c(x) = \sum_{\alpha} u_\alpha(x) e^{i k_\alpha x} M_{\alpha}(x) + \cdots
\end{align}
Here $M_{\alpha}(x)$ runs over the 20 $4 \pi$-monopoles, $u_{\alpha}(x)$ is a periodic function within the unit cell, and $k_\alpha$ is the lattice momentum of monopole $\alpha$.
The lattice symmetries, specifically the magnetic algebra and a possible $C_n$ symmetry, will place some interesting constraints on the $u_\alpha$ and $k_\alpha$. 
To find these constraints we must work out how the lattice symmetries act on the monopoles.
The microscopic lattice symmetries will act as a subgroup of the larger emergent (Lorentz$\times \SU(3)\times U(1)_{\textrm{flux}}$) symmetry.
Specifically, under a lattice symmetry $R$, a monopole will transform under the general form
\begin{equation}
R: M_{S,m,a}^\dag \rightarrow L_{S,m}(R) U_{a,b}(R) e^{i\theta(R)} M_{S,m,b}^\dag.
\end{equation}
The monopoles are labeled by their IR quantum numbers, namely Lorentz spin $S$, the spin index $m=-S, -S+1, \cdots, S$ and $\SU(3)$ flavor indices $a, b$.
The first term $L_{S,m}$ represents the Lorentz transformation, which may depend on the Lorentz spin and spin index.
The second term $U_{a,b}$ represents the SU$(3)$  transformation.
The last term $e^{i\theta(R)}$ represents the $U(1)_{\textrm{flux}}$ transformation, which  comes from a Dirac sea contribution~\cite{Alicea_vortexliquid2,HermeleKagome,Alicea_monopole}. 
Therefore, it is independent of the Lorentz spin and $\SU(3)$  index.

We will consider the lattice translations along two primitive directions of a lattice ($T_1$ and $T_2$) as well as the lattice rotation $C_n$.
We assume they embed into the Lorentz part as the usual Euclidean space group, 
\begin{equation}
L_{S,m} (T_i) = 1, \quad \quad L_{S,m}(C_n) = e^{i 2\pi m/n}.
\end{equation}
For the SU$(3)$ flavor rotation, since we only have SU$(3)$ singlet and adjoint monopoles, they are in the same SU$(3)$-reps as the bilinear masses discussed in Sec.~\ref{sc:CDW}.
Specifically, we have $U=1$ for the SU$(3)$ singlet monopole, and for the SU$(3)$ adjoint monopole we have 
\begin{equation}
U_{a,b}(T_1) = \delta_{a,b} e^{ik_{1,a}}, \quad \quad U_{a,b}(T_2) = \delta_{a,b} e^{ik_{2,a}},
\end{equation}
where the eight members of the adjoint representation are labeled by the ``momenta'' $(k_{1,a}, k_{2,a})= \{ (\frac{2 \pi n_1 }{3}, \frac{2 \pi n_2}{3}) \}$ with $(n_1, n_2) \neq 0$ and $a=1, 2, \cdots, 8$. 
The $C_n$ must act on these just like they would on momenta in the Brillouin zone. Specifically, for $C_4$ (square lattice) or $C_6$ (triangular lattice), and an appropriate ordering of $a = 1, \cdots 8$
\begin{equation}
U(C_4) =\left( \begin{matrix}
0 & 1 & 0 & 0 & 0 & 0 & 0 & 0 \\
0 & 0 & 1 & 0 & 0 & 0 & 0 & 0 \\
0 & 0 & 0 & 1 & 0 & 0 & 0 & 0 \\
1 & 0 & 0 & 0 & 0 & 0 & 0 & 0 \\
0 & 0 & 0 & 0 & 0 & 1 & 0 & 0  \\
0 & 0 & 0 & 0 & 0 & 0 & 1 & 0  \\
0 & 0 & 0 & 0 & 0 & 0 & 0 & 1 \\
0 & 0 & 0 & 0 & 1 & 0 & 0 & 0  \\
\end{matrix}
\right),
\end{equation}
and
\begin{equation}
U(C_6) =\left( \begin{matrix}
0 & 1 & 0 & 0 & 0 & 0 & 0 & 0 \\
0 & 0 & 1 & 0 & 0 & 0 & 0 & 0 \\
0 & 0 & 0 & 1 & 0 & 0 & 0 & 0 \\
0 & 0 & 0 & 0 & 1 & 0 & 0 & 0 \\
0 & 0 & 0 & 0 & 0 & 1 & 0 & 0  \\
1 & 0 & 0 & 0 & 0 & 0 & 0 & 0  \\
0 & 0 & 0 & 0 & 0 & 0 & 0 & 1 \\
0 & 0 & 0 & 0 & 0 & 0 & 1 & 0  \\
\end{matrix}
\right),
\end{equation}

At last, we need to work out the phase factor $\theta(R)$ from the U$(1)_\textrm{flux}$ rotation. 
Physically it comes from the contribution of Dirac sea, which can be determined numerically (e.g. see Ref.~\cite{Alicea_vortexliquid2,Alicea_monopole}).
Usually $\theta(R)$ can only take quantized values due to the symmetry constraint.
Specifically, on the square lattice we have
\begin{align}
T_2 C_4 &= C_4 T_1, \\
T_1 C_4 & = C_4 T_2^{-1}, \\
(C_4)^4 &= 1.
\end{align}
On the triangular lattice we have,
\begin{align}
C_6 T_2 & = T_1^{-1} C_6, \\
T_2 C_6 & = C_6 T_1 T_2, \\
(C_6)^6 & = 1.
\end{align}
The monopoles are local electron operators seeing integral flux, so they should satisfy the above algebraic relations.
Based on the transformation of the SU$(3)$ singlet monopole, we will have
\begin{equation}\label{eq:square_monopole}
\theta(T_i) = s\pi, \quad \theta(C_4) = \frac{(2n+1)\pi}{4}
\end{equation}
on the square lattice. 
Here $s$ and $n$ are integers that can be determined numerically.
Similarly on the triangular lattice, we have
\begin{equation}\label{eq:triangular_monopole}
\theta(T_i) = 0, \quad \theta(C_6) = \frac{(2n+1)\pi}{6}. 
\end{equation}

\subsubsection{Experimental consequences}
With the symmetry properties of the monopoles in hand, we are ready to discuss their experimental detection.
Since the electron can be expanded in monopole operators (See Eq.~\eqref{eq:cdecomp}), any  measurement  sensitive to the electron Green's function will probe the two-point function of the monopoles. For energy resolved spectroscopies, such as STM and ARPES, each monopole will contribute an energy dependence of the form $E^{2 \Delta_\alpha - 1}$, where $\Delta_\alpha$ is its scaling dimension. 
In our case, there are two such dimensions, corresponding to $\SU(3)$ singlet and adjoint monopoles, which may be quantitatively similar. 
So it would be particularly spectacular if symmetry could be used to impose selection rules whereby the two dimensions can be measured separately. 

We first consider a momentum resolved spectroscopy such as ARPES (or more relevant to heterostructures, planar tunneling  \cite{Jang2017}), which probes the spectral function $A(E, k)$. Following Eq.~\eqref{eq:cdecomp}, $A(E, k)$ will contain a singular component from each monopole with momentum $k = k_\alpha$, which  is just the action of $T_{1/2}$ on the monopole. On the triangular lattice,  the low-energy spectral weight will peak at $(k_1, k_2)= \{ (\frac{2 \pi n_1 }{3}, \frac{2 \pi n_2}{3}) \}$:
$(n_1, n_2)=(0,0)$ corresponds to the $S=3/2$, SU$(3)$ singlet monopoles, while the other 8 momenta correspond the the $S=1/2$, SU$(3)$ adjoint monopoles.
So momentum perfectly unravels the two exponents: $A(E, k = 0) \propto E^{2 \Delta_{S=3/2} - 1}$, while $A(E, k_\alpha \neq 0 ) \propto E^{2 \Delta_{S=1/2} - 1}$. 
On the square lattice, the results are similar except that the relevant momenta may have a shift of $(\pi, \pi)$ coming from the contribution $\theta(T_i)=\pi$ in Eq.~\eqref{eq:square_monopole}.
The 8 distinct momenta should again be degenerate, giving another way to detect the emergent SU$(3)$ symmetries.

The second option is to use STM spectroscopy. At a generic point, the STM will couple to both types of monopoles, so that $\frac{dI}{dV} \sim a  V^{2 \Delta_{3/2} - 1} + b  V^{2 \Delta_{1/2} - 1} + \cdots$.
However, if the STM is above a position with point-group symmetry (a ``Wyckoff position''), then the STM will only measure monopoles with zero lattice angular momentum about the point (these extinctions arise from the constraints $C_n$ imposes on the $u_{\alpha}(x)$ of Eq.~\eqref{eq:cdecomp}).
Combining the results of $L_{S,m}(C_n)$, $U(C_n)$ and $\theta(C_n)$, we find that the 16 SU$(3)$ adjoint monopoles (with $S=1/2$) will always contain both zero and finite angular momentum monopoles, regardless of whether we consider the triangular or square lattice, so their scaling dimension will always manifest in $\frac{dI}{dV}$.
The 4 SU$(3)$ singlet monopoles (with $S=3/2$), on the other hand, may or may not contain zero angular momentum monopoles - specifically, they will not on the triangular lattice if $n=2,3$ in Eq.~\eqref{eq:triangular_monopole}. This case would be particularly convenient as the $\Delta_{3/2}$ contribution to $\frac{dI}{dV}$ would  vanish above these high-symmetry points.

	In the presence of dilute impurities, STM can also be used to analyze the wavevectors present in the resulting Friedel oscillations. Analyzing this ``quasiparticle inference''\cite{WangLee2003} would be another way to obtain a $k$-resolved monopole spectroscopy.

\section{Confinement transition of the $1/2$ Laughlin state: application of  level-rank duality \label{sc:level-rank}}
Some of the critical theories in Eq.~\eqref{eq:critical} may enjoy a  further enhanced symmetry.
One example is the phase transition between the bosonic $\nu=1/2$ Laughlin state and a trivial insulating phase.  
This transition can be described by  QED$_3$-Chern-Simons theory with $C_1=0$, $C_2=1$ and $k=1$~\cite{Chen1993_transition, Maissam2014_FQHtransition},
\begin{equation}
\mathcal L= \bar \psi   (i\slashed{\partial} + \slashed{a}) \psi +\frac{3}{8\pi} a da -\frac{1}{2\pi} A da +\frac{1}{4\pi} A dA - m \bar \psi \psi.
\end{equation}
The critical theory has a bosonic dual~\cite{Seiberg2016395,karchtong}, which is~\cite{Wen1993_transition}
\begin{equation}
\mathcal L = |(\partial_\mu - i b_\mu) \phi|^2 -\frac{2}{4\pi} b d b +\frac{1}{2\pi} A db - m |\phi|^2 - u |\phi|^4.
\end{equation}
Here $\phi$ is a O$(2)$ scalar, and $b$ is a U$(1)$ dynamical gauge field.
Since $\phi$ itself carries an unit charge of the gauge field $b$, $\phi$ can be understood as the semionic quasiparticle of the $1/2$ Laughlin state. 
Tuning the mass term $m|\phi|^2$ gives two phases and transition: i) For the positive mass, the semions are gapped and we get the $1/2$ Laughlin state. ii) For the negative mass, the $\phi$ are condensed and the gauge field $b$ is Higgsed, giving rise to a trivial insulating state.  

At  face value, the above two critical theories only have a U$(1)$ symmetry, the flux conservation of the dynamical gauge field (microscopically, the number conservation of bosons).
However, it was conjectured that they may have an enlarged SO$(3)$ symmetry~\cite{Benini2017}.
It is based on the conjectured level-rank duality~\cite{seiberg2}, namely those two theories may be dual to a $\SU(2)$ Chern-Simons theory,
\begin{equation}
\mathcal L = \bar \Psi (i\slashed\partial + \slashed \alpha +\slashed{ A}/2) \Psi + \frac{1}{8\pi}(\textrm{CS}[\alpha]) +\frac{1}{16\pi} A d A - m \bar \Psi \Psi.
\end{equation}
Here we have a $\SU(2)$ fundamental Dirac fermion $\Psi$ coupled to a level-1/2 $\SU(2)$ Chern-Simons gauge field $\alpha_\mu$. $A_\mu$ is a U$(1)$ probing field.
$\textrm{CS}[\alpha]$ is a shorthand for $\epsilon_{\mu\nu\rho} \textrm{tr} \left( \alpha_\mu \partial_\nu \alpha_\rho+\frac{2}{3}\alpha_\mu\alpha_\nu\alpha_\rho\right)$, which is $4\pi$ times the $\SU(2)$ Chern-Simons term of $\alpha_\mu$ with level-1.
By tuning the mass of Dirac fermions, one would either get a $\SU(2)_1$ Chern-Simons theory or zero Chern-Simons: the former corresponds to the $1/2$ Laughlin state~\cite{Wen_nonAbelianFQH}
while the latter corresponds to a trivial insulating phase. 

All the three theories can describe the transition from the $1/2$ Laughlin state to a trivial insulator. 
Therefore, one would conjecture they are dual to each other.   
Furthermore, one would expect the first two theories have an enhanced symmetry (rather than a naive U$(1)$ symmetry), since the last one has a manifest SO$(3)$ symmetry --the Dirac fermion $\Psi$ forms a fundamental representation of both the $\SU(2)$ gauge symmetry and the SO$(3)$ global symmetry. 
This duality and symmetry enhancement were already conjectured in Ref.~\cite{Benini2017}, and here we take one step further to identify the relation among operators in three different theories,
\begin{align}
\bar \psi \psi & \sim |\phi|^2 \sim \bar \Psi \Psi, \label{eq:SO3_mass} \\
\nabla \times a & \sim \nabla \times b  \sim \bar \Psi \sigma_\mu \Psi \label{eq:density} \\
\mathcal M_a \psi^\dag & \sim \mathcal M_b (\phi^*)^2 \sim \Psi^T(\sigma_{\mu}\sigma_y\otimes\tau_y)\Psi \label{eq:pairing}
\end{align}
The $\mathcal M_a$ ($\mathcal M_b$) is the bare monopole that creates $2\pi$ flux of the gauge field $a_\mu$ ($b_\mu$). Due to the Chern-Simons term, one should attach $\psi^\dag$ ($(\phi^*)^2$) to the monopole in order to have a local (gauge neutral) object (for instance see Ref.~\cite{ChesterPufuChern-Simons}).  
$\sigma_{\mu=1,2,3}$ are Pauli matrices acting on the Dirac index, while $\tau_y$ acts on the gauge index.

The first line (Eq.~\eqref{eq:SO3_mass}) is the mass term that tunes through the phase transition, and is a scalar under the SO$(3)$ (and Lorentz) symmetry.
The second line identifies the flux in the U$(1)$ gauge theory with the gauge-invariant density and current of $\Psi$ fermions in the SU$(2)$ gauge theory.
In the third line, we have the monopole operators in the U$(1)$ gauge theory dual to the pairing amplitude and current of $\Psi$ fermions in the  SU$(2)$ gauge theory.
To understand the last equation, we should realize that due to the Chern-Simons term, the monopole operators are carrying Lorentz spin-$1$ (hence they have three complex components).
Moreover, $(\nabla \times a, \textrm{Re}(\mathcal M_a \psi^\dag), \textrm{Im}(\mathcal M_a \psi^\dag))$ forms a SO$(3)$ vector, and they are 
the conserved current of the conformal field theory, whose scaling dimension is precisely 2. 

We expect that the operators in Eq.~\eqref{eq:SO3_mass}-\eqref{eq:pairing} are the ones with lowest scaling dimension. 
Therefore, they will give the dominate contribution to the physical operator correlation functions. 
Since the boson creation (or density) operator has the same global quantum number as the monopole (or flux) operator, we will have
\begin{equation}
b_0^\dag b_r \sim 1/r^4, \quad \quad n_0 n_r \sim 1/r^4.
\end{equation}

Therefore, in a bosonic system with only particle conservation, the critical point would enjoy an enlarged SO$(3)$ symmetry that makes the boson creation operator to be degenerate with the boson density operator.
The above results may be tested in numerical simulations or experiments.
We remark that the SO$(3)$ symmetry is most transparent in the chiral spin liquid system~\cite{Kalmeyer1987,Wen1989,YCH14,Gong2014,Bauer2014} where it is nothing but the spin rotation symmetry (The $SU(2)$ gauge theory can be constructed from a $SU(2)$ parton construction~\cite{Wen_nonAbelianFQH}). 
In this context, the spin operator $\vec S$ has scaling dimension $2$, while the dimer operator $\vec S_0 \cdot \vec S_1$ corresponds to the mass term in Eq.~\eqref{eq:SO3_mass}.

\section{Experimental realization \label{sc:experiments}}

In this section we briefly describe the experimental setup for FCI transitions, and the scenario we believe is simplest to realize.
The required ingredients are Landau levels and a tunable periodic potential, which are already available in experiments. 
There is no particular type of tunability required, e.g., one could change the detailed form of the superlattice $V_m$, the Landau level spacing, or various pseudospin Zeeman fields, but conceptually it is simplest to tune the overall strength of the potential ($U_0$ in our model). Then i) for small potential strength, the system realizes a conventional FQH state that has electrons partially filling the lowest Landau level; ii) for large potential strength, the Landau level splits into sub-bands, and the electrons may partially fill a sub-band, yielding a FCI state.
A phase transition between the FQH and the FCI could then be expected at a critical potential strength, as found in our numerics.
An experimental setup with a tunable potential strength has been realized recently in Refs.~\cite{Dean2017,Yankowitz2017}, where a  range of $U_0 / E_c$ can be realized.

The existence of a transition can be diagnosed from the behavior of the gap in a Wannier plot, as in Fig.~\ref{fig:general_transition}. If two experimentally observed  $(C,s)$ trajectories intersect at $n_\ast, \phi_\ast$, one or both with fractional $C$, and the pair exchange stability as the potential strength is tuned, then QED$_3$ may describe the transition. 
To fall within our QED$_3$-Chern-Simons analysis, the two gaps should be attributable to states in the same Jain sequence (e.g., the same $k$), and for pure QED$_3$, the two states should be particle-hole conjugates  ($\sigma_{xy}=p/(2p+1)$--$\sigma_{xy}=(p+1)/(2p+1)$). We describe the $n_\ast, \phi_\ast$ where they might appear  in Appendix~\ref{app:pureQED3}.

\begin{figure}[t]
\includegraphics[width=0.49\textwidth]{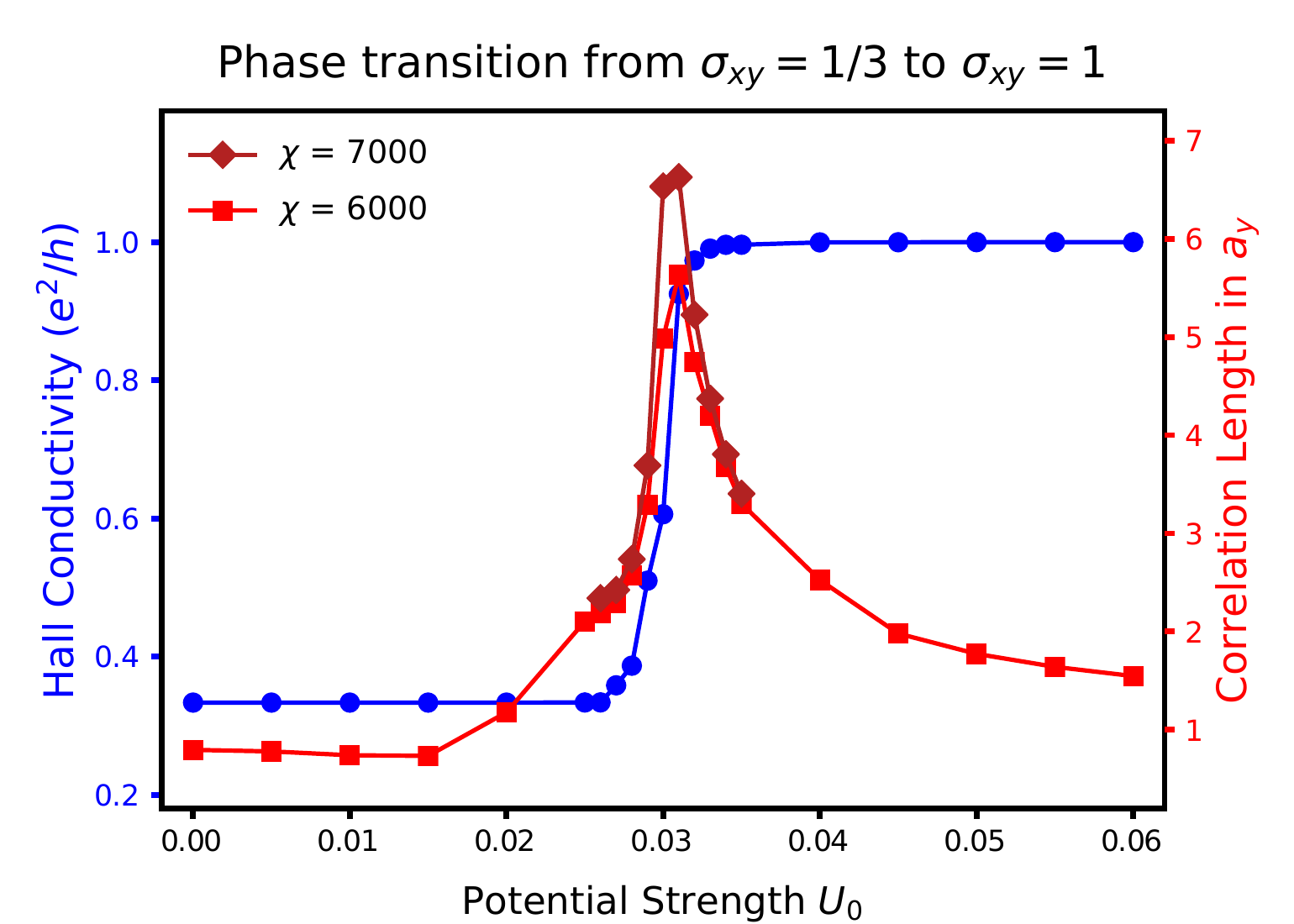}
\caption{\label{Phase_Transition2} The DMRG simulation result on infinite cylinder geometry with circumference size $L_y = 24.6 l_B$. $\phi =3/2$, $n = 1/2$, $V_1 = V_2 = 1$, $V_3 = V_4 = 1.6$ while changing the potential strength $U_0$. Blue line represents the change of Hall conductivity, which is measured at bond dimension $\chi=2000$. Red lines represent the change of correlation length with bond dimensions $\chi=6000,7000$. 
}
\end{figure}

While the numerical results we have presented thus far were for a FQH-FCI transition, experimentally a FQH-ICI (Integer Chern Insulator) transition may be a simpler initial target, since ICI states are more stable and easy to realize in experiment.
For example, consider a transition between $\sigma_{xy}=1/3$ FQH and $\sigma_{xy}=1$ ICI, which can occur at flux density $\phi_\ast=3/2$ and electron density $n_\ast=1/2$, with the same potential as in the FQH-FCI transition of Sec.~\ref{sc:FQH_FCI}.
For small $U_0$ the $\sigma_{xy}=1/3$ FQH state is realized.
For large $U_0$, the LLL splits into three sub-bands, and the electrons completely fill the lowest sub-band whose Chern number is $1$.
So $U_0 / E_C$ is expected to tune between the two, realizing  $N_f = 2, K = 1/2$ QED$_3$-Chern-Simons described further in Appendix~\ref{app:FQH_CI}.
We have also carried out iDMRG simulations to confirm the above scenario.
In Fig.~\ref{Phase_Transition2}, we observe a sharp transition across the value of $U_0 \sim 0.03$, where the Hall conductivity changes from $1/3$ to $1$ and the correlation length increases with bond dimension $\chi$. As in the case of the transition between $\sigma_{xy}=1/3$ FQH and $\sigma_{xy}=2/3$ FCI, numerical evidence supports a continuous transition.

\section{Summary and Discussion \label{sc:summary}}

Motivated by the recent experimental realization of FCIs in graphene heterostructures, we have shown how phase transitions between different FCIs induced by changing the lattice potential can be used to realize the whole family of QED$_3$-Chern-Simons theories.
A key ingredient is that the magnetic algebra of the microscopic model can be used to realize multiple symmetry related Dirac fermion flavors ($N_f > 1$) without further fine tuning.
At the critical point, this leads to an emergent $\SU(N_f)$ symmetry which can be diagnosed from the electron compressibility at finite momentum. 

We hope  our  work will stimulate further investigation of  QED$_3$-Chern-Simons theory.
For example, previous literature has mostly focused on the situation where $N_f/2 +K$ is an integer due to the parity anomaly~\cite{kapustinqed,ChesterPufuChern-Simons}.
Nevertheless, it has now been realized that this constraint can be relaxed: $N_f/2+K$ can be a half integer, or other fractions.
Mathematically, the parity anomaly can be avoided by either forbidding certain monopole operators~\cite{MaxAshvin15,wangsenthil15b},  by introducing an auxiliary topological Chern-Simons term~\cite{Seiberg2016395}, or  redefining the gauge charge of Dirac fermions~\cite{SeibergLargeCharge}.
This new sequences of ``pseudo-anomalous" QED$_3$-Chern-Simons theory have rarely been investigated~\cite{SeibergLargeCharge}.
The critical theories which are straightforward within the proposed experimental realization have $N_f/2+K$ half integer.
For example, the transition we studied in Sec. \ref{sc:FQH_FCI} (between the $\sigma_{xy}=1/3$ state and $\sigma_{xy}=2/3$ state) has $N_f=3$ and $K=0$.
It will be interesting to have a better theoretical understanding of these theories.

Another interesting direction is to understand the properties of monopoles in QED$_3$-Chern-Simons theory~\cite{kapustinqed,ChesterPufuChern-Simons,ChesterPufuBootstrappingQED,DyerMonopoleTaxonomy}.
As discussed, the monopole is nothing but the electron, so the scaling dimensions of the monopoles can be detected using standard spectroscopic techniques such as STM. 
On the other hand, there are also opportunities to explore those critical theories under the deformation of adding monopole terms.
For a given $N_f, K$ QED$_3$-Chern-Simons theory, if the monopole operators proliferate (condense), will the theory flow to a symmetry breaking state,  a topologically ordered state, or a new CFT~\cite{WangDCPdual}?
In the context of the FCI transition, monopole proliferation may be achieved by proximity to a superfluid or superconductor.

In the current work we have not addressed the effect of the long-range Coulomb interaction and disorder.
In graphene heterostructures, the long-range Coulomb tail is screened by the proximity to metallic gates, so can likely be ignored.
Disorder, however, will be present. While the underlying graphene, as well as moire superlattices, are extremely pristine, the most tunable superlattice architecture, gate-patterning, \cite{Dean2017} will likely prove much dirtier.
However, in contrast to the conventional quantum plateau transition, here it is the lattice potential rather than disorder which enables a direct transition between different plateaus.
Therefore, the disorder may be ignored when the temperature is larger than the effective scattering rate. 
Nevertheless it is physically interesting to consider the effect of Coulomb interactions and/or disorder.
These perturbations may be marginal in the UV, and become either marginally relevant or irrelevant in the deep IR.
Once the perturbations are relevant, the critical theory will lose conformal invariance and flow into another fixed point.
The new fixed point could either correspond to a gapped state or a metallic state.
The answer to these questions will depend on the details (i.e. $N_f$ and $K$), and our understanding is incomplete even in the large$-N_f$ limit~\cite{Ye1998_transition,disorderQED3_Goswami,ThomsonSachdev2017}. These questions are left for future work.

\acknowledgements
We thank B. I. Halperin, C. Dean, L. Iliesiu and A. Young for stimulating discussions.
The DMRG numerics were performed using code developed in collaboration with Roger Mong and the TenPy collaboration.
YCH was supported by the Gordon and Betty Moore Foundation under the EPiQS initiative, GBMF4306, at Harvard University. JYL and AV were supported by the ARO MURI on topological insulators, grant W911NF-12-1-0961 and by a Simons Investigator award. CW was supported by Harvard Society of Fellows.
This research was enabled in part by support provided by  Compute Canada (www.computecanada.ca) and Cedar.

\appendix

\section{Details of the large-$N_f$ calculation}\label{Appendix:LargeN}
The calculation will largely follow Ref.~\cite{Hermele2005,Hermele2005_Erratum}. The only modification is on the photon propagator due to the Chern-Simons term.

We assume large $N_f$ and $k$, with the ratio $\lambda=\frac{8k}{\pi N_f}$ fixed. The effective Euclidean Lagrangian of the the gauge field $a_{\mu}$, at leading order in $1/N_f$, is given by
\be
\mathcal{L}[a_{\mu}]=\frac{N_f}{16}|q||\vec{a}(q)|^2+\frac{k}{4\pi}\epsilon_{\rho\mu\nu}q_{\rho}a_{\mu}(q)a_{\nu}(-q)+\frac{N_f}{\xi|q|}(\vec{q}\cdot\vec{a})^2,
\ee
where the first term comes from the one-loop contribution from the Dirac fermions, the second term is the Chern-Simons term, and the last term is a (non-local) gauge-fixing term. The original Maxwell term is irrelevant and omitted.

We can now choose a convenient $\xi$ (Feynman gauge) to make the photon propagator of the following form:
\begin{fmffile}{P}
\begin{equation}
\begin{gathered}
\begin{fmfgraph*}(65,50)
\fmfleft{i1}
\fmfright{o1}
\fmf{photon}{i1,o1}
\end{fmfgraph*}
\end{gathered}=\frac{16}{N_f(1+\lambda^2)}\left[\frac{\delta_{\mu\nu}}{|q|}-\frac{\lambda}{q^2}q_{\rho}\epsilon_{\rho\mu\nu} \right].
\end{equation}
\end{fmffile}

The fermion line and the vertex are standard:
\begin{fmffile}{F}
\begin{equation}
\begin{gathered}
\begin{fmfgraph*}(65,50)
\fmfleft{i1}
\fmfright{o1}
\fmf{fermion}{i1,o1}
\end{fmfgraph*}
\end{gathered}=\frac{\slashed{k}}{k^2},
\end{equation}
\end{fmffile}
\begin{fmffile}{V}
\begin{equation}
\begin{gathered}
\begin{fmfgraph*}(65,50)
\fmfleft{i1}
\fmfright{o1,o2}
\fmf{photon}{i1,v}
\fmf{fermion}{v,o1}
\fmf{fermion}{o2,v}
\end{fmfgraph*}
\end{gathered}=-\gamma_{\mu}.
\end{equation}
\end{fmffile}

At order $O(1/N_f)$, fermion self-energy is given by
\begin{fmffile}{SE0}
\begin{equation}
\Sigma_0(k)=\begin{gathered}
\begin{fmfgraph*}(55,65)
\fmfleft{i1}
\fmfright{o1}
\fmf{fermion, label=$k+q$}{i1,o1}
\fmf{photon,left, label=$q$}{i1,o1}
\end{fmfgraph*}
\end{gathered}=\frac{8}{3\pi^2N_f(1+\lambda^2)}\slashed{k}\ln(|k|/\Lambda),
\end{equation}
\end{fmffile}
where only the logarithmically divergent part is kept and $\Lambda$ can be viewed as the UV cutoff (the exact meaning depends on regularization scheme and is not important).

Now consider turning on an $SU(N_f)$-singlet mass perturbation
\begin{fmffile}{M}
\begin{equation}
\begin{gathered}
\begin{fmfgraph*}(65,50)
\fmfleft{i1}
\fmfright{o1}
\fmfv{decor.shape=cross}{v}
\fmf{fermion}{i1,v,o1}
\end{fmfgraph*}
\end{gathered}=im.
\end{equation}
\end{fmffile}

The diagrams relevant for calculating its scaling dimension at $O(1/N_f)$ are
\begin{fmffile}{SE1}
\begin{equation}
\Sigma_1(k)=\begin{gathered}
\begin{fmfgraph*}(65,65)
\fmfleft{i1}
\fmfright{o1}
\fmfv{decor.shape=cross}{v}
\fmf{fermion}{i1,v,o1}
\fmf{photon,left}{i1,o1}
\end{fmfgraph*}
\end{gathered}=-\frac{24}{\pi^2N_f(1+\lambda^2)}(im)\ln(|k|/\Lambda),
\end{equation}
\end{fmffile}
and
\begin{fmffile}{SE2}
\begin{eqnarray}
\Sigma_2(k)&=&\begin{gathered}
\begin{fmfgraph*}(65,65)
\fmfleft{i1,i2}
\fmfright{o1,o2}
\fmfv{decor.shape=cross}{v}
\fmf{fermion}{i1,o1}
\fmf{photon}{i1,i2}
\fmf{photon}{o1,o2}
\fmf{fermion,left}{o2,i2}
\fmf{fermion}{i2,v,o2}
\end{fmfgraph*}
\end{gathered}+\begin{gathered}
\begin{fmfgraph*}(65,65)
\fmfleft{i1,i2}
\fmfright{o1,o2}
\fmfv{decor.shape=cross}{v}
\fmf{fermion}{i1,o1}
\fmf{photon}{i1,i2}
\fmf{photon}{o1,o2}
\fmf{fermion,right}{i2,o2}
\fmf{fermion}{o2,v,i2}
\end{fmfgraph*}
\end{gathered} \nonumber \\
&=&\frac{64(1-\lambda^2)}{\pi^2N_f(1+\lambda^2)^2}(im)\ln(|k|/\Lambda).
\end{eqnarray}
\end{fmffile}

The scaling dimension $\Delta_{m_s}$ can then be extracted following standard procedure:
\be
\Delta_{m_s}=2-\frac{64}{3\pi^2(1+\lambda^2)N_f}+\frac{64(1-\lambda^2)}{\pi^2(1+\lambda^2)^2N_f}+O(1/N_f^2).
\ee

For the $SU(N_f)$-adjoint mass, $\Sigma_1$ is identical but $\Sigma_2$ is absent due to the cancelation in the fermion loop. The scaling dimension is then
\be
\Delta_{m_a}=2-\frac{64}{3\pi^2(1+\lambda^2)N_f}+O(1/N_f^2).
\ee

\section{Microscopic setting for the transition}

In this section, we will describe the microscopic setting for the phase transition. 
The results are summarized in Table. \ref{table:realization}.

\begin{table}
\caption{\label{table:realization} The summary of interesting sequences of transitions. We consider a transition from $\nu=p/(2p+1)$ FQH state to a FCI or CI state. We take a simple square lattice potential in Eq.~\eqref{eq:potential}, and each unit cell has the flux density $\phi$ and the particle density $n$. The multi-flavor Dirac fermions are protected by the magnetic translation symmetry of composite fermions.} 
\setlength{\tabcolsep}{0.25cm}
\renewcommand{\arraystretch}{1.4}
\begin{tabular}{c|ccccc}
\hline\hline
  Transition to   & $N$ & $K$ & $\phi$ & $n$   \\ \hline
 $\sigma_{xy}=\frac{p+1}{2p+1}$ FCI & $2p+1$ & $0$ & $2p$ & $\frac{2p^2}{2p+1}$ \\ \hline
 $\sigma_{xy}=1$ ICI & $p+1$ & $p/2$ & $\frac{2p+1}{p+1}$ & $\frac{p}{p+1}$ \\ \hline
 $\sigma_{xy}=0$ ICI & $p$ & $(p+1)/2$ & $\frac{2p+1}{p}$ & $1$ \\ 

\hline \hline
\end{tabular}
\end{table}

\subsection{FQH-ICI transition \label{app:FQH_CI}}

Compared with the FCI, integer Chern insulator (ICI) is much easier and robust to realize in experiments.
Therefore, it is interesting to consider the transition from the $\nu=p/(2p+1)$ FQH phase to a ICI phase.
Since the ICI phases with $C=0,1$ also belong to the Jain sequence, they correspond to composite fermions form the $C_1=0$ or $C_1=-1$ integer state,
So the FQH-ICI transition can be described by our QED$_3$-Chern-Simons theory Eq.~\eqref{eq:critical}.
Specifically, the transition to the $C=0$ ICI has the critical theory,
\begin{align}
\mathcal L &= \sum_{I=1}^{p}  \bar \psi_I  (\slashed{\partial} - i \slashed{a}) \psi_I + \frac{p+1}{8\pi} a d a + \frac{1}{8\pi} A d A -\frac{1}{4\pi} a d A.
\end{align} 
Similarly, the transition to the $C=1$ ICI is described by,
\begin{align}
\mathcal L &= \sum_{I=1}^{p+1}  \bar \psi_I  (\slashed{\partial} - i \slashed{a}) \psi_I + \frac{p}{8\pi} a d a + \frac{1}{8\pi} A d A -\frac{1}{4\pi} a d A.
\end{align} 
We don't write the mass term $m \sum \bar \psi \psi$  explicitly, but we remark that due to the finite Chern-Simons term the mass $m$ should be tuned in order to hit the criticality.

Microscopically the FQH-ICI transition is indeed straightforward to realize. 
Again we consider the Landau levels under a weak superlattice potential.
For example, suppose we have a super-lattice with the flux density $\phi = (2p+1)/p$  and particle density $n=1$ in each unit cell.
When the potential is weak, the $\nu=p/(2p+1)$ FQH is naturally realized.
When the potential is strong compared with the Coulomb energy, the LLL splits into $2p+1$ sub-bands, and the $p$ lowest sub-bands are filled.
By a proper choice of potential pattern, it is not hard to realize that the $p$ lowest sub-bands has the total Chern number  $C=0$. 
Consequently we could imagine a transition from the $\nu=p/(2p+1)$ FQH to the $C=0$ ICI by tuning the potential strength. 
Furthermore the composite fermion sees an effective flux $\phicf = 1/p$, which means that the magnetic translation symmetry will protect $p$ Dirac cones at the critical point.

Similarly we can construct the model for the transition between the $\nu=p/(2p+1)$ FQH and the $C=1$ ICI state.
One way to realize it is to have the flux $\phi = (2p+1)/(p+1)$  and density  $n=p/(p+1)$ in each unit cell.
In the weak potential limit, we have the $\nu=p/(2p+1)$ FQH state.
In the strong potential limit, the Landau level splits into $2p+1$ sub-bands, and the $p$ lowest sub-bands are filled. 
The total Chern number of the lowest $p$ sub-bands can be $1$ under a proper choice of potential pattern.
Also it is easy to show that the $p+1$ Dirac cones at the critical point is protected by the magnetic translation symmetry since the effective flux of the composite fermion is $\phicf=1/(p+1)$.

Indeed a simple square lattice potential in Eq.~\eqref{eq:potential} gives precisely the required free fermion band structure, namely the lowest $p$ sub-bands has a total Chern number $0$ ($\phi=(2p+1)/p$) or $1$ ($\phi=(2p+1)/(p+1)$).

\subsection{Odd flavor QED3 transition \label{app:pureQED3}}

The pure QED$_3$ theory describes the transition between the particle-hole conjugate partner of the Jain sequence states. 
For the fermionic system, the pure QED$_3$ theory (Eq.~\eqref{eq:fermionic_QED3}) has the odd number ($2p+1$) of Dirac fermions, and the transition happens between the $\sigma_{xy}=p/(2p+1)$ and the  $\sigma_{xy}=(p+1)/(2p+1)$ states.
As we already discussed in Sec. \ref{sc:FQH_FCI} , the $1/3-2/3$ (i.e. $p=1$) transition can be realized in a simple setting. 
Here we generalize this setting to the arbitrary $p$.

We consider a superlattice potential that has the flux $\phi=2p$ and the density $n=2p^2/(2p+1)$ in each unit cell. 
The filling factor is $\nu=n/\phi=p/(2p+1)$, thus a FQH state is naturally expected when the potential is weak.
Once the potential is finite, the lowest Landau level splits into $2p$ bands. 
Under a square lattice potential (e.g. Eq.~\eqref{eq:potential} with $V_1=V_2=1$, $1\gg V_3=V_4>0$), the Chern number of the bands is: $C_I=0$ when $I=1, \cdots, p-1, p+1, \cdots 2p$, and $C_{p}=1$. 
The electrons completely fill the  $p-1$ lowest bands ($C=0$), and partially fill (with the filling factor  $(p+1)/(2p+1)$) the  $C=1$ band (the $p_{\textrm{th}}$ band).
Hence the $\sigma_{xy}=(p+1)/(2p+1)$ FCI may appear, and the $p/(2p+1)-(p+1)/(2p+1)$ transition could be realized by tuning the potential strength.
Furthermore, one can show that the composite fermion has an effective flux $\phicf=\phi-2n=2p/(2p+1)$, hence the $2p+1$ Dirac cones of the critical theory are protected by the magnetic translation symmetry.

To realize the above scenario microscopically, one may need to further adjust the potential pattern as well as the interaction. If the lattice potential is not adjusted appropriately, we may fall into a Fermi liquid phase or a CDW phase since these two phases can possibly compete with a FCI phase when the potential is non-zero.
As we discussed in Sec.~\ref{sc:FQH_FCI}, for the case with $p=1$, one needs to adjust the diagonal potential $V_3=V_4$ ($\approx 1.4 V_1$) to have a large flatness ratio in the $C=1$ band. The gapped $\sigma_{xy} = 2/3$ phase is only observed for the small range of $V_3/V_1$ near $1.4$, and outside of this range, a metallic phase was observed, signaling metal-FCI transition. We expect that similar tuning of the lattice potential would also be required for a larger $p$, whose systematic investigation is left for the future work.


\end{document}